\documentclass[superscriptaddress,aip,jcp,amsmath,amssymb,showkeys,floatfix,
reprint
]{revtex4-2}

\usepackage{bibentry}

\usepackage[framemethod=tikz]{mdframed}
\mdfsetup{
	outerlinewidth=1pt,
	linecolor=white!90!black,
	innertopmargin=6pt,
	innerbottommargin=6pt,
	leftmargin=2pt,
	rightmargin=2pt
}

\usepackage{color}

\usepackage{graphicx}
\usepackage{dcolumn}
\usepackage{bm}
\usepackage{xfrac}
\usepackage{colortbl}

\usepackage{natbib}
\usepackage{bibentry}
\usepackage{float}
\usepackage{sidecap}
\usepackage[section]{placeins}

\usepackage[caption=false]{subfig}

\usepackage{amsmath}
\usepackage{amssymb}
\usepackage{physics}
\usepackage{marvosym}
\usepackage{wasysym}
\usepackage{MnSymbol}
\usepackage{stmaryrd}
\usepackage{pifont}
\usepackage{wrapfig}
\usepackage{xifthen}
\usepackage{enumerate}
\usepackage{cancel}
\usepackage{mathtools}

\newcommand{\mbf}[1]{\mathbf{#1}}

\newcommand{\adn}{\hat a}
\newcommand{\aup}{\hat a^\dag}

\newcommand{\IG}[2][width=\linewidth]{\includegraphics[#1]{plots/#2}}

\renewcommand{\Ref}[1]{Ref.~\onlinecite{#1}}
\newcommand{\Refs}[1]{Refs.~\onlinecite{#1}}
\newcommand{\Fig}[1]{Fig.~\ref{f:#1}}
\newcommand{\Figure}[1]{Figure~\ref{f:#1}}
\newcommand{\Figs}[2]{Figs.~\ref{f:#1} and~\ref{f:#2}}

\newcommand{\Eq}[1]{Eq.~\eqref{e:#1}}

\newcommand{\Section}[1]{Section~\ref{s:#1}}

\newcommand{\etal}{\textit{et al}}

\newcommand{\sub}[1]{\textsubscript{#1}}

\newcommand{\Order}[1]{\mathcal{O}(#1)}

\usepackage[normalem]{ulem}

\newcommand{\processor}{Intel Core i7-11800H}
\newcommand{\laptop}{\processor\ laptop}

\usepackage{algorithm}
\usepackage{algpseudocodex}
\newenvironment{malgorithmic}[1][]{%
	\raggedright\begin{minipage}{0.9\linewidth}
		\begin{algorithmic}[#1]}{\end{algorithmic}
\end{minipage}\\[5pt]}

\begin{document}

\raggedbottom

\title[]{A truncated Davidson method for the efficient ``chemically accurate'' calculation of full configuration interaction wavefunctions without \emph{any} large matrix diagonalization}

\newcommand{\KBR}{KBR, 601 Jefferson St., Houston, TX 77002}

\newcommand{\QuAIL}{Quantum Artificial Intelligence Laboratory (QuAIL), NASA Ames Research Center, Moffett Field, CA 94035}

\newcommand{\ISD}{Intelligent Systems Division, NASA Ames Research Center, Moffett Field, CA 94035}

\author{Stephen J. Cotton}
\affiliation{\QuAIL}
\affiliation{Employed by \KBR}

\date{\today}

\begin{abstract}

	This work develops and illustrates a new method of calculating ``chemically accurate'' electronic wavefunctions (and energies) via a truncated full configuration interaction (CI) procedure which arguably circumvents the large matrix diagonalization that is the core problem of full CI and is also central to modern selective CI approaches.
	This is accomplished simply by following the standard/ubiquitous Davidson method in its ``direct'' form---wherein, in each iteration, the electronic Hamiltonian \emph{operator} is applied directly in second quantization to the Ritz vector/wavefunction from the prior iteration---except that (in this work) only a small portion of the resultant expansion vector is actually even computed (through application of only a similarly small portion of the Hamiltonian).
	Specifically, at each iteration of this truncated Davidson approach, the new expansion vector is taken to be twice as large as that from the prior iteration. 
	In this manner, a small set of highly truncated expansion vectors (say 10--30) of increasing precision is incrementally constructed, forming a small subspace within which diagonalization of the Hamiltonian yields clear, consistent, and monotonically variational convergence to the approximate full CI limit.
	The good efficiency in which convergence to the level of chemical accuracy (1.6~mHartree) is achieved suggests, at least for the demonstrated problem sizes---Hilbert spaces of $10^{18}$ and wavefunctions of $10^8$ determinants---that this truncated Davidson methodology can serve as a replacement of standard CI and complete-active space (CAS) approaches, in circumstances where only a few chemically-significant digits of accuracy are required and/or meaningful in view of ever-present basis set limitations.

\end{abstract}


\maketitle


\section{Introduction \& Background}\label{s:introduction}

Within \textit{ab initio} quantum chemistry, the method of configuration interaction (CI) stands as the most rigorous solution to the electronic structure problem---i.e.\ solving the electronic Schr\"odinger Eq.\ with the nuclei as fixed point charges, typically, in a basis of single particle basis functions (orbitals) determined through the Hartree-Fock self-consistent field (SCF) procedure. 
The cost of a full CI (FCI) calculation, however, scales exponentially or, more specifically, factorially with the number of ways of arranging the electrons amongst the given orbitals. It is therefore prohibitively expensive for all but the simplest molecular systems and basis sets.

Nevertheless, FCI remains important as a rigorous benchmark against which all other methods (all of which are approximate) may be judged. Numerical methods for FCI also serve as the core many-body solvers in complete active space CI (CASCI) approaches and orbital-optimizing complete active space self-consistent field (CASSCF) approaches, methods which are made more tractable by restricting consideration to a complete set of electronic excitations only within a strategically-selected set of chemically-relevant orbitals. The state-averaged variant of the latter (SA-CASSCF) finds particularly important use in calculations for excited electronic states, including the calculation of excitation energies, force gradients, and  non-adiabatic couplings needed for treating the electronically non-adiabatic dynamics of nuclear motion on multiple electronic potential energy surfaces (PESs). 
These important applications of CI, as well as FCI's status as the definitive solution in \textit{ab initio} quantum chemistry, make CI-based approaches obvious targets for quantum computing technology once the hardware has advanced sufficiently to allow these methods to be brought to bear on new interesting and/or important molecular systems. 

\newcommand{\bsigma}{\boldsymbol \sigma}
\newcommand{\bb}{\mbf b}
\newcommand{\bc}{\mbf c}

Of course, due to the simplicity and brute force rigour of FCI it has a many decades long history. Early on, it was appreciated that although the size of the FCI wavefunction scales factorially, it is actually quite sparse, meaning that the vast majority of the elements of the CI vector are nearly zero. 
Knowles,\cite{Knowles1989} and then Knowles with Handy,\cite{KnowlesHandy1989} seem to have been the first to explicitly leverage this fact in computations by employing a pair of numerical thresholds/cutoffs in the standard Davidson method,\cite{Davidson1975} specifically, to the subspace expansion/basis vectors $\{\bb_i\}$, and also to the result of applying the Hamiltonian matrix $\mbf H$ to each expansion vector, referred to as the $\bsigma$-vector in the FCI literature.\footnote{What is generally referred to as the $\bsigma$-vector is written as $\mbf g \equiv \mbf H \cdot \bc$ in \Refs{Knowles1989,KnowlesHandy1989}, with $\bc$ being an expansion vector.}
Validating the approach required repeating the entire Davidson procedure for variously decreasing pairs of cutoff values until reaching the desired convergence; ultimately, this showed that near-exact FCI energies could be obtained at a fraction of the cost of conventional FCI.

This work was followed several years later by  Mitrushenkov\cite{Mitrushenkov1994} who assumed the same basic strategy of leveraging the sparseness of the CI vector but, instead, specified a single threshold based on the magnitude of each component of the $\bsigma$-vector itself---instead of basing it on a perturbative estimate of energy contribution as Knowles and Handy had done---and applied the truncation in the context of Olson \etal\@.'s\cite{OlsenEtAl1988,OlsenEtAl1990} modified version of the Davidson method.
Importantly, in this ``dynamic CI'' method, as in the earlier method of Knowles and Handy, the \emph{full} $\bsigma_i$ is initially calculated \emph{exactly} in each iteration~$i$ and then used, prior to truncation, to compute elements of the subspace energy matrix,
$H_{ij} \equiv \bb_i \cdot \mbf H \cdot \bb_j \equiv \bb_i \cdot \bsigma_j$. These matrix elements must be calculated exactly---even if the basis has been truncated---to the extent it is desired to preserve the variational nature of the calculations. 
The approximation is therefore limited to truncation of the expansion vectors themselves---to the use of a truncated $\bsigma_i$ to generate the next expansion vector $\mbf b_{i+1}$---which, as noted in \Refs{Knowles1989,KnowlesHandy1989,Mitrushenkov1994,RolikEtAl2008}, should only affect the rate of convergence and number of iterations.

These same observations were made again many more years later by Rolik, Szabados, and Surj\'an,\cite{RolikEtAl2008} but with added focus on the key fact that calculation of the \mbox{$\bsigma$-vector} itself is the real computational bottleneck.
Their solution was to implement a so-called ``groping step'' which identified a subset of important determinants based on the prevalence of separate $\alpha$- and $\beta$-electron configurations occurring in the wavefunction of the current iteration, subject to filtering based on a second-order perturbative estimate of each determinant's contribution to the energy (so, reverting somewhat back to Knowles and Handy). 
From there, the method involved a ``correction step'' comprising a single Davidson iteration in which the Hamiltonian was applied in the subspace identified in the groping step, thereby avoiding full $\bsigma$-vector construction.

Interestingly, the so-called ``groping step'' of Rolik \etal.\@ is very reminiscent of the determinant selection step of selective CI (SCI) approaches, which represent---along with, e.g., FCI quantum Monte Carlo (FCIQMC)\cite{BoothEtAl2009,ClelandEtAl2010,ClelandEtAl2012} and density matrix renormalization group (DMRG)\cite{White1993,WhiteMartin1999,MitrushenkovEtAl2001,ChanHead-Gordon2002,Olivares-AmayaEtAl2015} methods---one of a few currently dominant paradigms for performing approximate and/or near-exact FCI calculations. 
Although SCI approaches initially appeared\cite{HuronEtAl1973} almost 50 years ago with the method of CI by perturbative selective iteration (CIPSI),\cite{HuronEtAl1973, EvangelistiEtAl1983} there has been a resurgence of interest\cite{Eriksen2021} in recent years, in part, spurred by the intervening development and success of FCIQMC, which has resulted in various different flavors of SCI---for instance, the original CIPSI, but also stochastic heat-bath CI (SHCI),\cite{HolmesTubmanEtAl2016a, SharmaHolmesEtAl2017a, HolmesUmrigarEtAl2017, LiOttenEtAl2018} adaptive CI (ACI),\cite{SchriberEvangelista2016} adaptive sampling CI (ASCI),\cite{TubmanLeeEtAl2016}, iterative CI with selection (iCI)\cite{LiuHoffmann2016,ZhangEtAl2020} (a relative of the previous iterative CI (ICI)\cite{Nakatsuji2000,NakatsujiDavidson2001}), and probably many other variations.

The crux of SCI is performing first the aforementioned selection step of choosing important determinants, followed by a second step of forming the matrix of the Hamiltonian in this space of selected determinants and diagonalizing it, typically using a standard approach (such as, again, the Davidson method). These steps, determinant selection followed by large-matrix diagonalization, are then repeated until the resulting SCI energy (and/or wavefunction) converges to within some threshold, though that convergence is not necessarily monotonic. 
SCI methods, therefore, perform many steps of explicit large matrix formation and diagonalization, though the matrix is orders of magnitude smaller than the FCI matrix.
This is an important distinction between SCI and the work of Rolik \etal.\@ along with the work of this paper: here there is no explicit large matrix diagonalization. 

The various flavors of SCI differ in the specifics of their determinant selection and also in how they deal with the obvious question of deciding, \textit{a priori}, how many determinants to choose. 
Also, in order to calculate FCI energies to chemical accuracy while keeping the dimensionality of the SCI matrix to a workable size, it is
typically the case that the variational solution to the large matrix problem is supplemented with some standard flavor of perturbation theory (e.g., this is the stochastic aspect of SHCI).
This enables the calculation of very accurate ground state energies for very large Hilbert spaces but, again, it may seem somewhat ambiguous how one decides, \textit{a priori}, how to divide Hilbert space into variational and perturbative treatments,
and the perturbative aspect may limit application in CAS-type methods for determining excited states, which are some of the most important practical application of these FCI-like approaches. An excellent review of these and other approaches to the FCI problem is found in \Ref{FalesEtAl2018}. See also \Ref{Eriksen2021} for a short and recent comprehensive ``perspective'' on the present and future of FCI. 

\begin{figure}
    \centering
    \IG{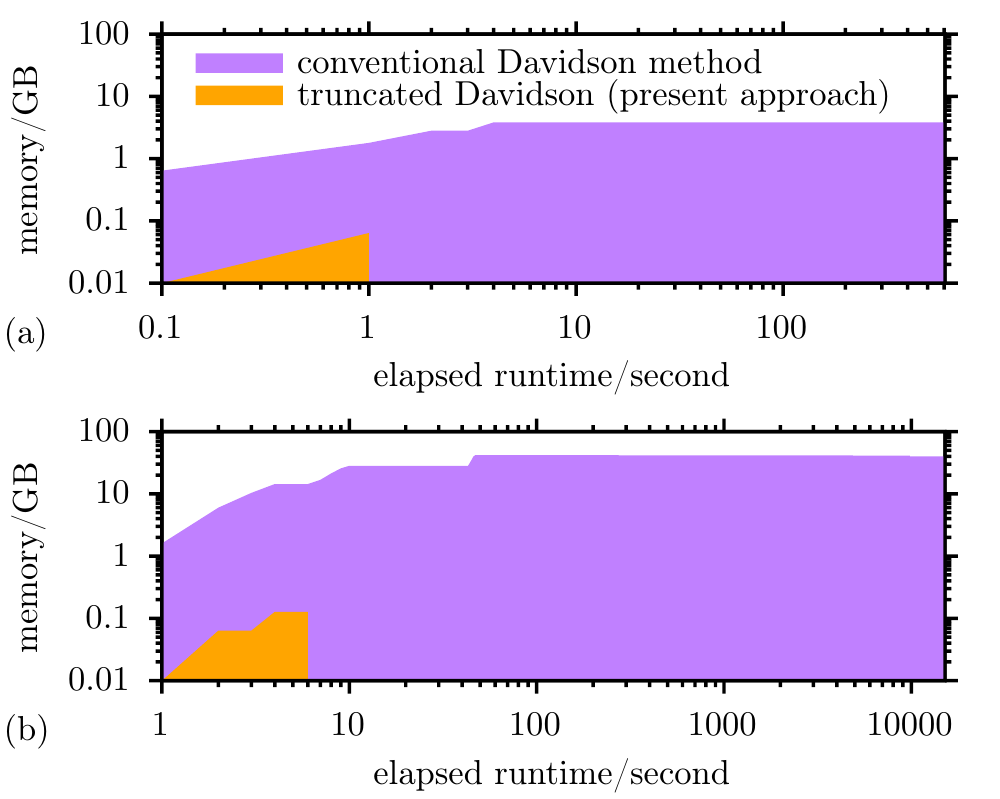}
    \caption{FCI using the conventional Davidson method compared with the present truncated Davidson approach: memory usage versus runtime to chemical accuracy for H\sub2O on a consumer-grade \laptop\ processor with 64GB of memory. (a) 6-311G basis (19 orbitals, none frozen, $r = 0.9455 \text{\AA}$, $\theta = 111.881 \text{\textdegree}$), and (b) 6-311G* basis (24 orbitals, none frozen, $r = 0.9394 \text{\AA}$ $\theta = 107.5\text{\textdegree}$), both as treated by Rolik \etal\@. in \Ref{RolikEtAl2008}. The present truncated Davidson approach uses around $1/10$ of a gigabyte (GB) of memory and converges to chemical accuracy in about 1~second in (a) and in about 5~seconds in (b). In contrast, the standard Davidson method in (b) requires around $42$~GB of memory and over 4~hours of processing time to reach chemical accuracy}

    \label{f:H2O,timings}
\end{figure}

The work presented in this paper is a purely variational approach (though it could, in principle, be supplemented with perturbation theory) which avoids SCI's hallmark steps of explicit determinant selection followed by large-matrix diagonalization.
Instead, what is developed here is essentially a highly-truncated version of the standard Davidson method, implemented in terms of sparse basis/expansion vectors subject to progressively decreasing numerical cutoffs. 
Retrospectively, this work may thus be viewed as an outgrowth of the early approaches of  Knowles, Handy, and Mitrushenkov; but, it goes further, in the spirit of Rolik \etal.\@, to the extent that full $\bsigma$-vector formation is carefully avoided, while preserving the variational structure of the calculations. It also adapts an important feature of SHCI to the construction of the expansion vectors in the sense that, of the second-quantized electronic Hamiltonian's quartic number of terms, only those deemed significant are ever even accessed in a given Davidson iteration.

The efficiencies resulting from such a straightforward truncation of the standard Davidson method as presented herein are significant. \Fig{H2O,timings} previews treating the classic example of water as considered\cite{RolikEtAl2008} by Rolik~\etal: panel~(a) shows treatment in the 6-311G basis set (of 19 orbitals) and panel~(b), treatment in the 6-311G* basis (of 24 orbitals). The figure shows that if one is interested in no more than chemical accuracy, this is achieved with the present approach in about 1~second for~(a) and in about 5~seconds for~(b), using about $1/10$ of a gigabyte (GB) of memory. This contrasts with the conventional Davidson method which requires hundreds of seconds for~(a) and over 4~hours of processing time for~(b), and moreover, for the later, using more than 42~GB of memory due to having to hold in memory the full $\bsigma$-vectors. Of course, other methodologies in this art demonstrate remarkable improvements over straightforward FCI; this preview illustrates the potential scale of these improvements and, in this case, the tiny memory footprint that can be achieved relative to conventional FCI.

The remainder of this paper is organized as follows: \Section{theory} explains the methodology, giving both rationalizing perspective and implementation details. \Section{results} presents some benchmark results, justifies the truncation parameters, and, again, contrasts the approach with the modern SCI paradigm. Conclusions and directions for future work are given in \Section{conclusion}.

\section{Theory \& Methodology}\label{s:theory}

The definitive many-electron problem of FCI is  
difficult (and a key \textit{raison d'\^etre} for quantum computers) because it represents a large matrix eigenproblem where the dimensionality of the matrix ($N$) scales factorially with the number of degrees of freedom ($n$), typically, in quantum chemistry, taken to be the occupancy of the molecular orbitals determined from an underlying Hartree-Fock (HF) self-consistent field (SCF) calculation. 

The FCI matrix is, however, extremely sparse---the vast majority of the elements are zero---and, moreover, what are sought are only a few of the lowest energy electronic states. Thus, the FCI eigenproblem is ideally suited for iterative sparse-matrix approaches which target only a few solutions at the low spectral end, such as the venerable Lanczos method\cite{Lanczos1950} as well as other, more robust and/or efficient, Krylov/power space-based approaches. 
The core numerical operation of these iterative sparse-matrix solvers is the application of the matrix to-be-diagonalized to various $N$-dimensional vectors, depending on the particular method employed. Because the $N \times N$ matrix is sparse, only the non-zero elements are stored (along with their indices) but, even stored in sparse format, the matrix still has $N$ diagonal elements (which are typically significant/non-zero). Moreover, the relevant $N$-dimensional CI-vectors, e.g., the eigenvectors sought, are generally stored in dense sequential (i.e., non-sparse) format and, again, for the FCI problem, the dimension $N$ is factorially-large.

As explained in \Section{introduction}, however, a key fact that may be leveraged to improve the situation is that the low-energy eigenstates of the FCI matrix are \emph{also} sparse; and, also: one does not necessarily seek a high-level of numeric precision, seeing as only a few digits, at best, carry any true physical significance in the typical quantum chemistry application where a vastly incomplete basis set is typically employed.

The point of this paper is, therefore, to directly leverage the sparseness of the CI vector by employing the notion of an inner-product space whose linear algebraic operations work on sparse vectors---lists of non-zero elements and their indices---subject to carefully chosen numeric cutoffs. It is easy to write reasonable algorithms for addition, scalar multiplication, and the inner product which work on these lists that are a tiny fraction of the size $N$ of the original dense vectors.
As for the FCI matrix itself which, although sparse, retains at least the factorial scaling of its $N$ diagonal elements, it may be replaced by the electronic Hamiltonian in second-quantized operator form,
\begin{equation}\label{e:H}
	\hat H = \sum_{i,a}^n \aup_a t_{ai} \adn_i + \sum_{i<j,a,b}^n \aup_a \aup_b u_{abij} \adn_j \adn_i,
\end{equation}
with exhibits only quartic scaling in $n$, the number electronic degrees of freedom, e.g., orbitals. The operator form of $\hat H$ (\Eq{H}) may be applied directly\cite{Roos1972} to the sparse vectors to generate the equivalent (though here approximate) of what would be the matrix-vector product, which, as stated, is all that is required by numerical methods for solving large sparse eigenproblems. This leaves no explicit dependence on the factorially-large $N$. 

The remaining question, in developing a method of coarse precision, is to pick a standard method of numerical linear algebra and to effectively translate it so as to work in an inner product space of sparse vectors, and then to carefully and aggressively choose truncation parameters for the sparse vectors to give the highest accuracy at the lowest cost. The most obvious choice is the ubiquitous Davidson method as reviewed next in \Section{Davidson}. A modified/truncated version of the Davidson method is then presented in \Section{truncated Davidson}. Justifications for the truncation scheme and numerical cutoffs will be given in the context of the results for various systems presented in \Section{results}.

\subsection{Review of the Davidson method}\label{s:Davidson}

\newcommand{\br}{\mbf r}
\newcommand{\bpsi}{\boldsymbol \psi}
\newcommand{\bpsiin}{\bpsi_\text{init}}
\newcommand{\bpsir}{\bpsi_\text{Ritz}}
\newcommand{\ER}{E_\text{Ritz}}
\newcommand{\ES}{E_\text{save}}
\newcommand{\eps}{\epsilon}

\begin{figure}

\begin{malgorithmic}[1]
\Function{Davidson}{$\hat H,\bpsiin,\eps$} \Comment{\scriptsize Given Hamiltonian operator, guess wavefunction, and energy tolerance}

	\State $\bb_1 \gets \bpsiin$ \Comment{first basis/expansion vector}

	\For{$n=1,2,3,\ldots$}

		\State Compute \label{sigma-formation}
		$\bsigma_n \gets \hat H\, \bb_n $. \Comment{direct method\cite{Roos1972}}

		\State Update subspace energy matrix, $ \mbf H: \{ H_{ij} \gets \bsigma_i \cdot \bb_j, \; \forall i,j \leq n \}$.

		\State Solve the subspace eigenproblem,
		\[
			\mbf H \, \bc = \ER \, \bc,
		\]
		for the lowest energy  solution.

		\State Compute the Ritz vector 
		\Comment{wavefunction}
		\[
			\bpsir \gets \sum_{i=1}^n c_i \, \bb_i .
		\]

		\If{ $n>1 \land \left| \ES - \ER \right| < \eps$ } \Comment{done} \State \Return $\ER,\bpsir$
		\Else
		\State $\ES \gets \ER$
		\Comment{save to check convergence}

		\LComment{Steps to compute the next basis vector}

		\State Compute the residual $[\hat H - \ER] \, \bpsir$,
		\[
			\br \gets \sum_{i=1}^n c_i\, (\bsigma_i - \ER\,\bb_i )
		\]

		\State Apply the diagonal preconditioner,
		$\br \gets \br / ( \ER - \hat D )$.

	\State Orthogonalize: $ \br \gets \br - \sum_{i=1}^n (\br \cdot \bb_i) \, \bb_i $
		\State Normalize: $ \br \gets \br /\sqrt{ \br \cdot \br } $
		\State Update the basis: $\{\bb_i\} \gets \br \cup \{\bb_i\}$
		\EndIf
	\EndFor

\EndFunction

\end{malgorithmic}

\caption{Pseudocode for the standard Davidson method using energy change as the stop condition. In each loop iteration, the Hamiltonian is applied one time, directly as an operator at line~\ref{sigma-formation}, to generate the $\bsigma$-vector, which is the most memory and processor-time intensive operation. If, after solving the subspace eigenvalue problem the energy is not converged, an additional basis vector for the next iteration is calculated as the preconditioned residual (orthonormalized to the prior basis vectors) using the ``diagonal'' operator $\hat D \equiv \sum_i t_{ii} \hat n_i  + \sum_{i<j} u_{ijij} \hat n_i \hat n_j$ where $n_i \equiv \aup_i \adn_i$}.

\label{f:Davidson}

\end{figure}

The standard Davidson method is outlined in \Fig{Davidson}. Starting with a given initial guess wavefunction (line 2)---say the Hartree-Fock solution or some linear combination of determinants, as perhaps given by CI singles---the method proceeds through a sequence of iterations (line 3) building a low-dimensional subspace of basis/expansion vectors~$\{\bb_i\}$ within which the matrix of the Hamiltonian is formed (line 5) and diagonalized (line 6). This gives the optimal solution to the eigenvalue problem within the span of the current basis vectors---the so-called Ritz value ($\ER$) and Ritz vector ($\bpsir$)---as computed from the expansion vectors (line 7) at the current iteration. At this point, the solution is checked~(line~8) and the procedure terminated~(line~9) if $\ER$ has changed by less than $\eps$, in this version.

In the first iteration, the expansion vector~$\bb_1$ is just the input guess state~$\bpsi$ (line 2). In subsequent iterations, the expansion vectors are calculated from the residual vector (line 13) corresponding to the current Ritz pair ($\ER,\bpsir$), subject to ``diagonal'' preconditioning~(line~14),
\begin{equation} \label{e:PCR}
	\frac 1{\hat D - \ER} [\hat H - \ER]\, \bpsir,
\end{equation}
where
\begin{equation}\label{e:D}
	\hat D \equiv \sum_i t_{ii} \hat n_i  + \sum_{i<j} u_{ijij} \hat n_i \hat n_j, \quad n_i \equiv \aup_i \adn_i.
\end{equation}
Note that although the preconditioner in \Eq{D} contains only the terms of $\hat H$ which do not change the state vector (the orbital occupations given by the determinants), its application is critical to accelerating the convergence of the Davidson procedure.
The preconditioned residual (PCR) vector, \Eq{PCR}, is then usually orthogonalized (line~15) to the current set of basis vectors $\{\bb_i\}$ and normalized (line~16) before appending it to the $\{\bb_i\}$ (line~17) for the next iteration.

The Hamiltonian is applied once per iteration, here, in \Fig{Davidson}, as an operator~$\hat H$ directly\cite{Roos1972}~(line~4), to the basis vector $\bb_i$ from the prior iteration. The result, $\bsigma_i$, is used both to construct the subspace energy matrix (line~5) as well as to compute the residual vector (line~13) needed for the next basis vector, $\bb_{i+1}$. As noted in \Section{introduction}, computation of the $\bsigma$-vector is the most computationally time- and memory-intensive operation in the Davidson procedure. See, also, \Ref{ParrishHohensteinEtAl2016} for another recent description of the canonical Davidson method in the quantum chemistry context.

\subsection{A truncated Davidson method using sparse vectors}\label{s:truncated Davidson}

\begin{figure}
\begin{malgorithmic}[1]

	\Function{truncated Davidson}{$\hat H,\bpsiin,\eps$}
	\Comment{\scriptsize Given Hamiltonian operator, guess wavefunction, energy tolerance}

	\State $\bb_1 \gets \bpsiin$
	\Comment{First basis/expansion vector}

	\For{$n=1,2,3,\ldots$}

	\State Update subspace energy matrix: \label{energy matrix construction}
	\Comment{avoiding $\bsigma$}

	\Statex \( \mbf H: \{ H_{ij} \gets \bb_i \cdot \hat H\, \bb_j, \; \forall i,j \leq n \} \)

	\State Update subspace overlap matrix: 

	\Statex \(\mbf S: \{ S_{ij} \gets \bb_i \cdot \bb_j, \; \forall i,j \leq n \}\)

	\State Solve the \emph{generalized} subspace eigenproblem, \label{eigenproblem}
	\[
		\mbf H\, \bc = \ER\, \mbf S\, \bc,
	\]
	for the lowest energy solution

	\State Compute the Ritz vector: \Comment{wavefunction}
	\[
		\bpsir \gets \sum_{i=1}^n c_i \, \bb_i .
	\] 

		\If{ $n>1 \land \left| \ES - \ER \right| < \eps$ } \Comment{done}
		\State \Return $\ER,\bpsir$
		\Else
		\State $\ES \gets \ER$
		\Comment{Save to check convergence}

		\LComment{Steps to compute the next basis vector}

		\State $T \gets 2 \times \text{\Call{sizeof}{$\bb_n$}}$ \Comment{Truncation target size} \label{expansion}

		\State $\delta \gets 1/T \in [\delta_\text{max},\delta_\text{min}]$  \Comment{Numerical threshold}

		where, e.g., $\delta_\text{max} = 10^{-3}$, $\delta_\text{min} = 10^{-8}$ 

		\State $\br \gets$ \Call{calc trunc'd PCR}{$\hat H, \bpsir, \ER, \delta$} \label{TPCR}

		\Comment{See \Fig{trunc'd PCR}}

		\State Further truncate $\br$ to target size $T$

		\State Normalize: $ \br \gets \br /\sqrt{ \br \cdot \br } $

		\State Update the basis: $\{\bb_i\} \gets \br \cup \{\bb_i\}$

		\EndIf
	\EndFor

\EndFunction

\end{malgorithmic}

\caption{The Davidson method of \Fig{Davidson} modified to avoid $\bsigma$-vector formation (see line~\ref{energy matrix construction}) and, moreover, to compute (in line~\ref{TPCR}) an aggressively truncated preconditioned residual (PCR) for use as the next basis/expansion vector in the Davidson procedure. Boldfaced variables are vectors stored in sparse format and updated with sparse vector operations. At each iteration, the basis/expansion vector $\bb_n$ expands by exactly a factor of~2 (line~\ref{expansion}) versus the size of $\bb_{n-1}$. Other changes include solving a generalized eigenvalue problem in line~\ref{eigenproblem} due to not orthogonalizing the $\{\bb_i\}$. Calculation of the truncated PCR (line~\ref{TPCR}) is described in detail in \Fig{trunc'd PCR} and the sparse vector operations used throughout are described in \Fig{sparse vector space}.}

\label{f:truncated Davidson}
\end{figure}

The main point of this work is to show how the standard Davidson method of \Fig{Davidson} can be translated to work with sparse vectors and to provide an efficient truncation scheme. \Figure{truncated Davidson} provides such a truncated Davidson algorithm and it is immediately seen that not a great deal has been altered versus \Fig{Davidson}, at least superficially.
The primary differences arise in construction of the truncated preconditioned residual (PCR) to be used as the next basis/expansion vector.

\paragraph{PCR expansion vector doubling.} The procedure for calculating a truncated PCR begins with the setting of a truncation target size~$T$ (line 13) for the next basis/expansion vector, as well as the setting of a related numerical threshold~$\delta$~(line 14).  
As shown, $T$ is set to be exactly twice that of the size of the current expansion vector~$\bb_i$. This means that if one chooses~$\bpsi_\text{init}$ to be the Hartree-Fock solution---the usual choice, which is a single determinant---the first expansion vector $\bb_1$ has a single element, the second~$\bb_2$ then only has 2~elements, the third~$\bb_3$, only 4~elements, and so forth. While eventually a size of~$2^{n-1}$ for $\bb_n$ becomes quite large, it will obviously never be anywhere near the size of the full Hilbert space if chemical accuracy (1.6 mHartree) is achieved after a reasonable number of Davidson iterations. What is remarkable is that chemical accuracy \emph{is} actually achieved in a reasonable number of iterations, as shown by the results in \Section{results}.

Allowing the basis vectors to double in size at each iteration follows the simple intuition that the action of $\hat H$ works to spread the wavefunction into Hilbert space and that a balanced, even-handed approach might allow as many new determinants as there are old/current ones. The numerics in \Section{results}, however, show that a factor of 2 turns out to be about the most efficient, although the success of the algorithm is not particularly sensitive to this specific choice so long as it is not too much different. I.e., choices of 3 or $\frac 32$ also work, just not quite as efficiently. Large growth factors, however---i.e., from not aggressively truncating---are very inefficient, essentially becoming equivalent to those of the standard Davidson method. 

Strictly requiring each expansion vector to be twice the size of that generated in the preceding iteration requires skipping the orthogonalization step normally present in the Davidson procedure, as shown in \Fig{Davidson} (line~15).  
As noted by Knowles,\cite{Knowles1989} orthogonalization would require the current expansion vector to contain every determinant present in those prior. While orthogonalization does not turn out not to be catastrophic to the present implementation, it is somewhat less efficient and does not seem to provide any noticeable benefit. 
Accordingly, the procedure shown in \Fig{truncated Davidson} skips the orthogonalization step and instead formulates a \emph{generalized} small-matrix eigenvalue problem (line~6), the solution of which poses no difficulty whatsoever.

\newcommand{\single}{\!\!\lcurvearrowup\!\!\protect\substack{a \\ i }\!\!\rcurvearrowup}
\newcommand{\double}{\!\!\lcurvearrowup\!\!\protect\substack{a b \\ i j}\!\! \rcurvearrowup}

\newcommand{\x}{c_{I'}}

\begin{figure}
	\begin{malgorithmic}[1]
		\Function{calc trunc'd PCR}{$\hat H,\bpsi,E,\delta$}
		\Comment{\scriptsize Given Hamiltonian operator, wavefunction, energy, threshold}

		\State $\br \gets \bpsi$ \Comment{From setting $\hat H = \hat D$ in \Eq{PCR}} 

	\LComment{Loop over terms of input state $\bpsi$}

	\For{$c_I,\ket{d_I}$ from $\bpsi = \sum_I c_I \ket{d_I}$} \label{psi}

		\LComment{1-electron moves via $[\hat D - E]^{-1}\hat H$}
		\For{ orbital $i$ occupied, $a$ unoccupied in $\ket{d_I}$}

		\State Let $ \ket{d_{I'}} \gets \ket{d_I \single}$ \Comment{ electron moved: $a \mapsfrom i$ }

		\State Compute $ \x \gets \bra{d_{I'}} \hat H \ket{d_I} \times c_I $

			\State Apply preconditioner: $\x \gets \x/( E_{d_{I'}}-E )$

			where $E_{d_{I'}} \ket{d_{I'}} = \hat D \ket{d_{I'}}$

			\If{ $|\x| > \delta$ } \Comment{avoids full $\bsigma$ formation}
				\State $\br \gets \br + \x \ket{d_{I'}}$
			\EndIf

		\EndFor

		\LComment{2-electron moves via $[\hat D - E]^{-1}\hat H$}
		\For{orbitals $i,j$ occupied in $\ket{d_I}$, $i < j$}
			\For{ orbitals $a,b$ unoccupied in $\ket{d_I}$, $a \neq b$}

				\State Let $ \ket{d_{I'}} \gets \ket{d_I \double}$ \Comment{ two electrons moved: $a,b \mapsfrom i,j$ }

				\State Compute $ \x \gets \bra{d_{I'}} \hat H \ket{d_I} \times c_I \propto u_{abij}\; c_I$

				\Comment{$\{u_{abij}\}$ grouped by $i,j$, each group sorted by magnitude}

				\If{ $|\x| < \delta$ }
					\State \textbf{break} $a,b$-loop \Comment{short-circuit to $i,j$-loop because $\{u_{abij}\}$ kept sorted} \label{short-circuit}
				\EndIf

				\vspace{5pt}

				\State Apply preconditioner: $\x \gets \x/( E_{d_{I'}}-E )$

					where $E_{d_{I'}} \ket{d_{I'}} = \hat D \ket{d_{I'}}$

				\If{ $|\x| > \delta$ } \Comment{avoids full $\bsigma$} 

					\State $\br \gets \br + \x \ket{d_{I'}}$

				\EndIf
			\EndFor
		\EndFor
	\EndFor

	\State \Return{$\br$}

	\EndFunction

	\end{malgorithmic}
	\caption{Pseudocode for the calculation of the diagonally-preconditioned residual (PCR), $[\hat D-E]^{-1}[\hat H - E]\, \bpsi$ (\Eq{PCR}), subject to the threshold $\delta$.
	The ``diagonal'' part of $\hat H$ replicates in the residual $\br$ the original terms of the input wavefunction $\bpsi$ (line~2), which is seen by setting $\hat H = \hat D$ in \Eq{PCR}.
	The contribution to $\br$ from the off-``diagonal'' terms of $\hat H$ is computed by looping over the $\{\ket{d_I}\}$ in  $\bpsi$ (line~4) and, for each $\ket{d_I}$, weighted by $c_I$, considering contributions from determinants which are 1-electron move away, $\ket{d_I \single }$ (lines~6--11), or 2-electron moves away, $\ket{d_I \double }$ (lines~13--21).
	For both 1- and 2-electron moves, the threshold~$\delta$ restricts the contributions which are added to the growing $\br$ (lines~10 and~20, respectively) to avoid what would otherwise effectively be formation of the full $\bsigma$-vector.
	The dominant expense is the double-loop over 2-electron moves whose cost is controlled by short-circuiting (line~18) the inner-loop over the unoccupied orbitals once the $\delta$-condition (line~17) is satisfied. 
	This assumes the $\{u_{abij}\}$ are stored in groups of common $i,j$, each sorted by magnitude over $a,b$.}
	\label{f:trunc'd PCR}
\end{figure}

\paragraph{Numerical thresholding.}
The actual calculation of the truncated PCR used \Fig{truncated Davidson} is given in the separate function shown in \Fig{trunc'd PCR}. As a practical matter, this requires the input of a numerical threshold~$\delta$ so that compute-time and memory are not significantly wasted calculating elements of the next basis vector which will ultimately be discarded, because not within the target truncation size~$T$. For simplicity, $\delta$ is chosen in \Fig{truncated Davidson} (line~14) to be inversely proportional to $T$, subject to a reasonable least inclusive bound of about chemical accuracy $10^{-3}$ (early iterations) and a most inclusive bound of say $10^{-8}$ (final iterations). The reasonableness of these bounds is verified by the numerics in \Section{results}.

For a given input $\delta$ and guess wavefunction $\bpsi$ with energy $E$---i.e., in \Fig{truncated Davidson}, the Ritz pair, $\bpsir$ and $\ER$---the function shown in \Figure{trunc'd PCR} 
begins by initializing the to-be-computed PCR vector~$\br$ with the input $\bpsi$. This is because replacing $\hat H$ with just its ``diagonal'' elements $\hat D$ in the PCR expression of \Eq{PCR} just gives the identify operation on $\bpsi$. I.e., initializing $\br$ with the input wavefunction~$\bpsi$ completely accounts for the terms of $\hat H$ which do not change orbital occupations as well as the residual energy subtraction in \Eq{PCR}.

The remainder of \Fig{trunc'd PCR}'s pseudocode concerns the effect of the ``off-diagonal'' terms of $\hat H$ in the application of the operator $[\hat D - E]^{-1} \hat H$ of \Eq{PCR}. $\hat H$ is, of course, the standard second-quantized electronic Hamiltonian operator, as shown in \Eq{H}, with $\aup_i$ and $\adn_i$ being fermionic creation and annihilation operators for the $i$th molecular orbital, and $\{t_{ai}\}$ and $\{u_{abij}\}$ being the standard 1- and 2-electron integrals over combinations of orbitals formed from the chosen basis set.
$\hat H$, being a 2-body operator,
its ``off-diagonal'' terms, i.e. those  which \emph{do} change the configuration of orbital occupations, generate candidate determinants which are either 1- or 2-electron moves away from determinants appearing in the input state $\bpsi$. Combinatorially, the 2-electron moves are by far the dominant expense, potentially greatly expanding the size of the wavefunction. 

Accordingly, as shown in \Fig{trunc'd PCR}, the computation of 1-~and 2-electron moves proceeds via a loop (line~4) over the expansion of the input state $\bpsi$,
\begin{equation}\label{e:Ritz}
	\bpsi = \sum_I c_I \ket{d_I},
\end{equation}
where $\ket{d_I}$ represents the $I$th particular configuration of orbital occupations---i.e., determinant---with amplitude/weight $c_I$ in $\bpsi$.
With respect to each determinant~$\ket{d_I}$, potential 1-electron moves are assessed first (lines~6--11): For the state $\ket{d_{I'}} \equiv \ket{d_I\single}$---where an electron has been moved from occupied orbital~$i$ to unoccupied orbital~$a$ (line~7)---its potential amplitude $c_{I'}$ in the growing residual vector $\br$ is the starting amplitude $c_I$ times the matrix element $\bra{d_{I'}} \hat H \ket{d_I}$ (line~8), subject to diagonal preconditioning (line~9), and, if $c_{I'}$ exceeds the $\delta$-threshold (line~10), $\ket{d_{I'}}$ is added to $\br$ (line~11) weighted by $c_{I'}$.
This restricting of the growth of $\br$ to only those contributions exceeding the $\delta$-threshold---roughly relating/corresponding to the target size~$T$ in the current iteration of the Davidson procedure of \Fig{truncated Davidson}---is the mechanism of controlling the cost of the algorithm and avoiding full $\bsigma$-vector formation.
Note that calculation of the matrix element $\bra{d_{I'}} \hat H \ket{d_I}$ (line~8) implicates multiple terms of $\hat H$ because both 1- and 2-body terms can move just a single electron. 

Next, potential 2-electron moves are considered (lines~13--21).
For the state $\ket{d_{I'}} \equiv \ket{d_I\double}$---where electrons have been moved from occupied orbitals~$i,j$ to unoccupied orbital~$a,b$ (line~15)---operations analogous to the case of 1-electron moves are performed: matrix element evaluation (line~16), diagonal preconditioning (line~19), $\delta$-thresholding (line~20), and conditional addition to $\br$ (line~21); except that, here, for 2-electron moves, 
there is an intervening threshold test (line~17) of the amplitude $c_{I'}$ being computed prior to preconditioning which provides an opportunity to short-circuit the innermost loop over the target orbitals $a,b$ greatly increasing the efficiency of the entire procedure, as now explained.

\paragraph{Abbreviated Hamiltonian access.} Still referring to \Fig{trunc'd PCR}, the consideration of 2-electron moves generated by $\hat H$ (lines~13--21) is the dominant computational expense in evaluation of the PCR.
However, compared to the 1-electron case, for a 2-electron move, computation of 
the matrix element $\bra{d_{I'}} \hat H \ket{d_I}$ only implicates a single term of $\hat H$, specifically,  that which is proportional to the single electron repulsion integral $u_{abij}$, as indicated in \Fig{trunc'd PCR} (line~16). Therefore, for a given $\delta$-threshold and amplitude $c_I$ of the starting state $\bpsi$, simple inspection of the table of $\{u_{abij}\}$ may be used to estimate the target amplitude $c_{I'}$ of $\ket{d_{I'}} \equiv \ket{d_I\double}$ before the preconditioning step and, moreover, so long as the electron repulsion integrals $\{u_{abij}\}$ are grouped by common $i,j$ and within each group sorted by magnitude over $a,b$ (which is trivial to do), the innermost loop over $a,b$-indices may be broken (line~18), once the $\delta$-threshold condition (line~17) is satisfied.

Neglecting the preconditioning step for the purposes of short-circuiting the innermost loop is a bit of an approximation, but it has been found to be a reasonable one numerically, both here and in 
\Ref{HolmesTubmanEtAl2016a}. It is precisely analogous to what is exploited in the heat-bath CI algorithm of \Refs{HolmesTubmanEtAl2016a,SharmaHolmesEtAl2017a,LiOttenEtAl2018}. Critically, this means that for most terms of $\bpsi$, whose coefficients $c_I$ are small, the vast majority of the quartic number of terms in the Hamiltonian operator $\hat H$ are \emph{not even accessed} during computation of the truncated PCR.
Of course, even prior to breaking the innermost loop, amplitudes post-preconditioning (line~19) must still satisfy the $\delta$-threshold (line~20) before corresponding determinants may be added to the growing PCR $\br$ (line~21). 

\newcommand{\ba}{\mbf a}
\newcommand{\bu}{\mbf u}
\newcommand{\bv}{\mbf v}
\newcommand{\bw}{\mbf w}

\begin{figure}
	\begin{malgorithmic}[1]

	\Function{multiply}{$s,\bv$} \Comment{\scriptsize given scalar $s$ and sparse vector $\bv$ whose elements are pairs of (\Call{index}{},\Call{value}{})}

	\State $\bu \gets \varnothing$

	\For{ $i=1,2,3,\ldots,\Call{sizeof}{\bv}$ }
		\State $u_i \gets v_i$
		\State $\Call{value}{u_i} \gets s \times \Call{value}{u_i}$
	\EndFor

	\State \Return $\bu$

	\EndFunction

	\vspace{12pt} 

	\Function{add}{$\bv,\bw$}
	\Comment{\scriptsize given sparse vectors $\bv$ and $\bw$ whose elements are pairs of (\Call{index}{},\Call{value}{})}

	\State $\bu \gets \bv \cup \bw$

	\State \Call{sort by index}{$\bu$} \Comment{Like states are now adjacent}

	\State $i \gets 1$

	\For{ $j=2,3,\ldots,\Call{sizeof}{\bu}$ }

		\If{ $\Call{index}{u_j} = \Call{index}{u_i}$ }

			\If{ $j \neq i$ }
				\State $u_i \gets u_i + \Call{value}{u_j}$
			\EndIf

		\Else

			\State $i \gets i+1$
			\State $u_i \gets u_j$

		\EndIf

	\EndFor

	\State \Call{erase}{$u_{i+1},u_{i+2},\ldots$}

	\State \Return{$\bu$}

	\EndFunction

	\vspace{12pt} 

	\Function{dot}{$\bv,\bw$} \Comment{\scriptsize given sparse vectors $\bv$ and $\bw$ whose elements are pairs of (\Call{index}{},\Call{value}{}) }

	\State \Call{sort by index}{$\bv$}
	\State \Call{sort by index}{$\bw$}
	\State $d \gets 0$
	\State $i \gets j \gets 1$
	\While{ $i \neq \Call{sizeof}{\bv} \wedge j \neq \Call{sizeof}{\bw}$ }

	\If{ $\Call{index}{v_i} < \Call{index}{w_j}$ }

		\State $i \gets i+1$

	\ElsIf{ $\Call{index}{v_i} > \Call{index}{w_j}$ }

		\State $j \gets j+1$

	\Else

	\State $d \gets d + \Call{value}{v_i}
					\times \Call{value}{w_j}$

	\EndIf

	\EndWhile

	\State \Return $d$

	\EndFunction

	\end{malgorithmic}
	\caption{Pseudocode for inner product space algebraic operations on sparse vectors used in \Figs{truncated Davidson}{trunc'd PCR}:
	$s\, \bv \equiv \text{\textsc{multiply}}(s,\bv)$,
	$\bv + \bw \equiv \text{\textsc{add}}(\bv,\bw)$,
	$\bv \cdot \bw \equiv \text{\textsc{dot}}(\bv,\bw)$; for a sparse vector $\bv$'s $i$th element, \textsc{index}($v_i$) accesses/returns its state/determinant index and, \textsc{value}($v_i$), the amplitude associated with that determinant}
\label{f:sparse vector space}
\end{figure}

\paragraph{Sparse vector implementation.}
Practical realization of the truncation-related efficiency benefits just described requires that the vectors, boldfaced in \Figs{truncated Davidson}{trunc'd PCR}, be given a sparse implementation. This means the vectors are represented as lists of elements, each element consisting of a value and an index, 
each index representing the occupation of the orbitals (i.e., a bit-string representing a determinant). \Figure{sparse vector space} provides self-explanatory pseudocode implementations for the sparse vector operations used in this work, these being just the standard algebraic operations on an inner-product space: scalar multiplication, addition, and the dot-product.\footnote{It is interesting to consider whether any other methods of numerical linear algebra could possibly be gainfully translated into a sparse vector implementation with the subroutines in \Fig{sparse vector space} or similar.}

\paragraph{Subspace energy matrix construction.}

The subspace energy matrix $\mbf H$ of the procedure in \Fig{truncated Davidson} is small---it is a $n \times n$ matrix at the $n$th Davidson iteration---and it is therefore trivial to solve the generalized eigenvalue problem (line~6) using a standard dense-matrix $\Order{n^3}$-scaling approach---in this work, a prior step of singular value decomposition (SVD) subject to a cutoff in the singular values of $10^{-7}$ (relative to the largest) was employed to deal with the non-orthogonality of the underlying basis $\{\bb_i\}$.

The \emph{construction} of $\mbf H$ (line~4), however, is a non-trivial task and one that is desirably accomplished \emph{exactly} in order to strictly preserve the variational nature of the Davidson procedure.
The difficulty is that, despite the low dimension of $\mbf H$, each basis vector is itself a linear combination of many determinants, i.e., $\bb = \sum_{I} c_I \ket{d_I}$, and so the computation of each element of $\mbf H$,
\begin{equation}\label{e:H expanded}\begin{aligned}
	H_{ij} &\equiv \bb_i \cdot \hat H\, \bb_j \\
	       &= \sum_{IJ} c_{iI} c_{jJ} \bra{d_I} \hat H \ket{d_J},
\end{aligned}\end{equation}
still involves finding all the ways the determinants in $\bb_j$ connect, through action of the Hamiltonian operator $\hat H$, to the determinants in $\bb_i$.
And, again, because of the aggressive truncation and unlike the standard Davidson method (\Fig{Davidson}), there is no $\bsigma$-vector (in \Fig{truncated Davidson}) which may give $\mbf H$ through a simple dot-product operation.

Nevertheless, there already exist well-developed methods in the SCI literature for building the (sparse) matrix of $\hat H$ in the space of the large number of selected determinants, and these same techniques can be utilized here for efficiently finding and evaluating the non-zero elements $\bra{d_I} \hat H \ket{d_J}$ in \Eq{H expanded}, except that, here, instead of storing all these connections individually, they can be immediately combined according to \Eq{H expanded} for all $H_{ij}$, simultaneously. This is particularly convenient to do if an implementation simply maintains a list of all relevant determinants along with the basis vector(s) in which each determinant occurs and the coefficient amplitude(s).
Furthermore, an implementation may incrementally construct $\mbf H$ one column (or row) at a time per Davidson iteration ($\mbf H$ being symmetric), and so at iteration $n$ the relevant connections are restricted to being those between the subset of determinants in $\bb_n$ and the union of the full set of determinants in $\{\bb_1,\ldots,\bb_n\}$.

In practice, if these steps are followed and state-of-the-art techniques in the spirit of those from the SCI literature are adapted, constructing 
the (small) subspace energy matrix $\mbf H$ via \Eq{H expanded} is typically comparable cost-wise to the other significant expense in \Fig{truncated Davidson}, which is the calculation of the truncated PCR (line~15) via the separate function described in \Fig{trunc'd PCR}.
In this work, the methods employed for constructing $\mbf H$ are very similar to (or at least inspired by) those utilized in the heat-bath CI (HCI) literature\cite{LiOttenEtAl2018} which, as stated, is arguably the most well-developed of modern SCI approaches. 
Again, in reference to \Fig{truncated Davidson}, the combined expense of updating $\mbf H$ (line~4), updating $\mbf S$ (line~5), and solving the generalized eigenproblem (line~6) is typically less than the calculation of the truncated PCR (line 15). Thus, while one may consider there to be factor-of-2-expense in the truncated approach of \Fig{truncated Davidson} due to applying $\hat H$ in two steps (lines 4 and 15), versus the standard Davidson method of~\Fig{Davidson} where it is is applied only once~(line~4), this key alteration enables a many orders of magnitude gain in efficiency by avoiding \Fig{Davidson}'s explicit computation of the exact $\bsigma$-vector~(line~4).

\section{results \& discussion}\label{s:results}

The results presented here are intended to demonstrate: (1) the basic viability and efficiency of the present approach through application to various modest-sized benchmark problems; (2) the smooth monotonic convergence one obtains to chemical accuracy and the robustness of the methodology at early iterations; (3) the rationale behind the choices of various parameters, such as the doubling of truncation target size $T$ at each iteration and the related selection of the $\delta$-threshold parameter; and lastly (4) the efficiency of the approach in comparison to state-of-the-art selective CI (SCI) approaches. 

For all the calculations presented here,
self-consistent field (SCF) optimized molecular orbitals at the restricted Hartree-Fock (RHF) level were generated---without utilizing any symmetry---via   
the \verb!Psi4! quantum chemistry software package,\cite{SmithEtAl2020}  except, for the Cr\sub 2 example, where they were generated with \verb!PySCF!.\cite{SunEtAl2018} The 1-~and 2-body electronic integrals over these orbitals---used in \Eq{H}---were written to ``disk'' (i.e., non-volatile memory) to be read in by a separate stand-alone code written by myself for the present work.

\subsection{H\sub 2O, a classic simple example}\label{s:H2O}

\begin{figure}
    \centering
    \IG{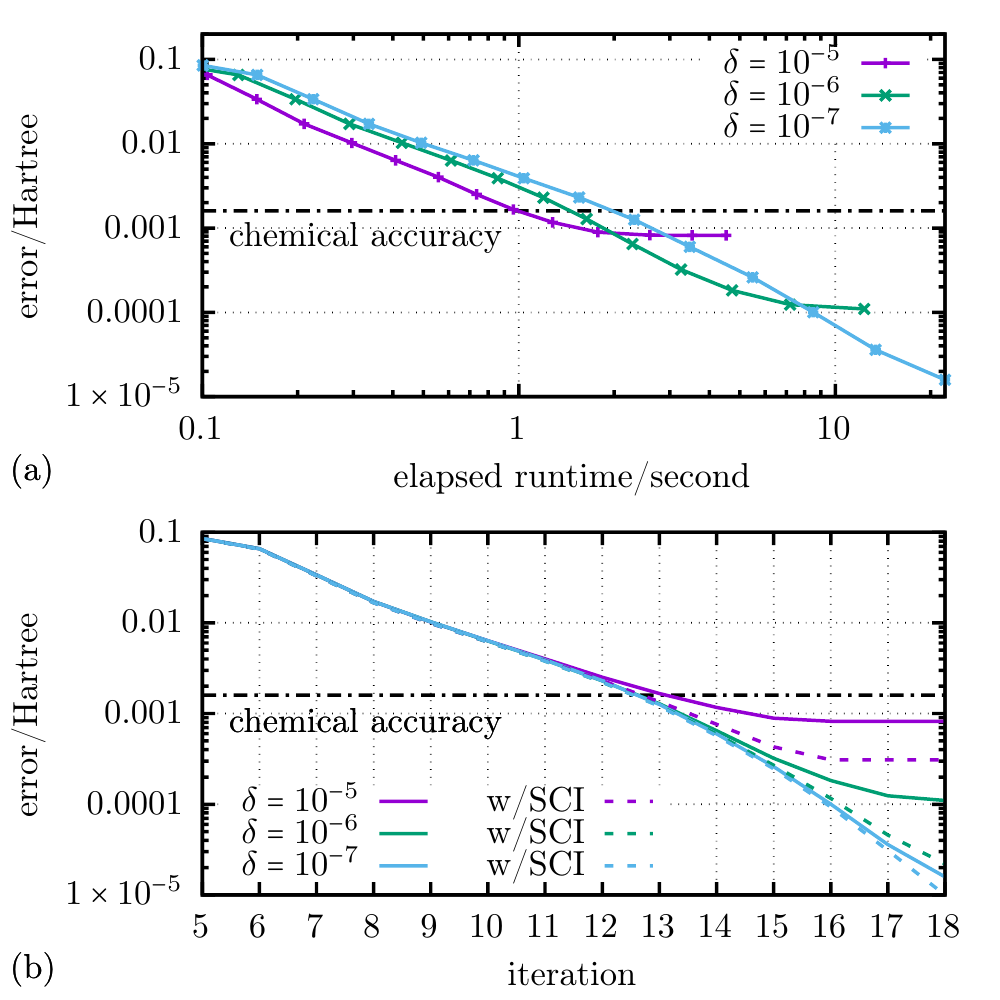}
    \caption{Convergence of the ground state energy of H\sub2O in the 6-311G basis (19 orbitals, none frozen) as computed with the present truncated Davidson approach at the Hartree–Fock optimized geometry, $r = 0.9455 \text{\AA}$, $\theta = 111.881 \text{\textdegree}$ as given in \Ref{RolikEtAl2008}. The error, relative to the exact FCI energy of~$-76.174823$ Hartree (\Ref{RolikEtAl2008}), is plotted on a log-scale for three different $\delta$-thresholds, as described in the text. (a) Error versus elapsed runtime (\laptop); (b) Error versus Davidson iteration number along with corresponding results---dashed lines with matching colors---generated by a SCI-type diagonalization in the determinant space corresponding to the union of all determinants in the expansion vectors}
    \label{f:H2O}
\end{figure}

The water molecule is chosen as a first illustrative example because of its frequent appearance in the early FCI literature and also because it was one of the examples treated by 
Rolik, Szabados, and Surj\'an in \Ref{RolikEtAl2008}, 
which is arguably the most pedagogically similar prior art.
\Ref{RolikEtAl2008} treated H\sub 2O in the modest 6-311G basis set of 19 orbitals at the Hartree–Fock optimized geometry $r = 0.9455 \text{\AA}$, $\theta = 111.881 \text{\textdegree}$ having a benchmark FCI energy of $-76.174823$ Hartree.

FCI results for H\sub 2O calculated with the present truncated Davidson approach (using the same geometry and 6-11G basis) are shown for comparison in \Fig{H2O}. Panel~(a) plots the error---i.e., the deviation from the exact FCI result---as a function of runtime for three different choices of the $\delta$-threshold parameter, as described in reference to \Figs{truncated Davidson}{trunc'd PCR}. For all three choices, the figure shows smooth monotonic convergence to chemical accuracy, and then to an order-of-magnitude or two beyond chemical accuracy for the smaller $\delta$s.

Of course, the main focus of this work is on doing the minimal amount possible to just achieve chemical accuracy and panel~(a) of \Fig{H2O} illustrates that this is indeed accomplished in~\emph{1 second} for this system. These calculations were performed on an \laptop\ processor with 8 cores running at a base clock of 2.3~GHz.
In comparison, the calculation for H\sub 2O from \Ref{RolikEtAl2008} took around \emph{450 seconds}.
This was in 2008 on a server-grade AMD Opteron processor with a single core probably running at \~2.4~GHz. Given the similar clock speeds, the timing difference is obviously not attributable solely to the different processors, despite having 8 cores in the newer model.

Likewise, 
\Fig{H2O,timings}
has already shown the timings and memory footprint 
of a state-of-the-art, but conventional, FCI calculation for this system (H\sub2O, 19 orbitals). It was run on the same consumer-grade Intel Core i7 model using the \verb!Psi4! quantum chemistry program,\cite{SmithEtAl2020} giving a runtime of about 210 seconds per iteration. With conventional FCI, however, chemical accuracy is reached in only 3 iterations, so at about \emph{630 seconds} of runtime, see \Fig{H2O,timings}, this is probably comparable to \Ref{RolikEtAl2008}'s method but still several orders of magnitude more expensive than the truncated Davidson method presented here, even when, for this simple system, memory-constraints are not an issue. 

Furthermore, also already shown in \Fig{H2O,timings} is a more costly calculation in the 6-311G* basis of 24 orbitals which, with conventional Davidson-based FCI (\verb!Psi4!), took over 4~hours to reach chemical accuracy, as shown in panel~(b), versus only about \emph{5~seconds} with the present truncated approach. For this system, Rolik \etal.\@ reported their fastest runtime of about \emph{4700 seconds},\footnote{This is to $1/10$ of chemical accuracy, but it is still the fastest calculation reported in Table~I of \Ref{RolikEtAl2008}.} so although perhaps somewhat faster than conventional FCI, it is still not comparable to the efficiency of the 5-second calculation shown in \Fig{H2O,timings}(b), at least for obtaining minimum chemical accuracy.

The other important point from \Fig{H2O}(a) concerns the effects of the $\delta$-threshold. Panel~(a) shows that a less inclusive $\delta$ results in faster convergence, e.g., to chemical accuracy for $\delta=10^{-5}$, but that improvements in accuracy level-off at some point regardless of the number of iterations, whereas more-inclusive $\delta$'s postpone this leveling-off until higher accuracies are achieved, even down to errors near $10^{-5}$ Hartree for $\delta=10^{-7}$. Reassuringly, a consistent trend is seen with the choice of $\delta$, and furthermore, chemical accuracy is reached in just a few seconds for each of these $\delta$s. 

\Figure{H2O}(b) explores the question of how much energy may be missing at each step by virtue of not performing a large-matrix diagonalization, as would be done after the selection step of SCI approaches. Here the relevant determinants may be taken as the union of all those appearing in the Davidson expansion vectors $\{\bb_i\}$. 
The solid lines in \Fig{H2O}(b) re-plot the results of panel~(a) versus Davidson iteration number and the dashed-curves having the same colors show the energy obtained at each iteration if $\hat H$ is diagonalized over the union of the determinants. The dashed-curves are always below the solid ones, as they must be since both calculations are variational and the set $\{\bb_i\}$ only constitutes a small subspace within the union of determinants. It is also reassuring to see that there is very little missing energy until after chemical accuracy is reached and that the onset of the energy gap roughly coincides with the leveling-off seen in panel~(a).

\newcommand{\bis}{\{\bb_i\}}

The most significant point of \Fig{H2O}(b), however, is illustrated by the calculations employing the most-inclusive threshold of $\delta=10^{-7}$. It is important to understand that the length of the expansion vectors does not depend on $\delta$ since the truncation size $T$ is always just double that of the previous vector. A more inclusive $\delta$ only implies that each $\bb_i$ is calculated, in a sense, more ``accurately,'' for a given truncation $T$. What is remarkable, then, is that the $\delta=10^{-7}$ curves in \Fig{H2O}(b) illustrate---having very little missing energy down to an accuracy of $10^{-5}$ Hartree---that if highly truncated $\bis$ are calculated sufficiently accurately, the linear combination of determinants forming each $\bb_i$ is nearly optimal, because no further optimization in the full space of determinants appreciably improves the energy. In the final iteration of the $\delta=10^{-7}$ calculation, this is a vector space of only 18 $\bb_i$ within a space of 454,988 determinants! In other words, no large-matrix diagonalization is required or even seemingly very beneficial in achieving chemically accurate FCI energies, so long as one calculates truncated Davidson expansion vectors with sufficient precision. 

\begin{figure*}
    \centering
    \IG{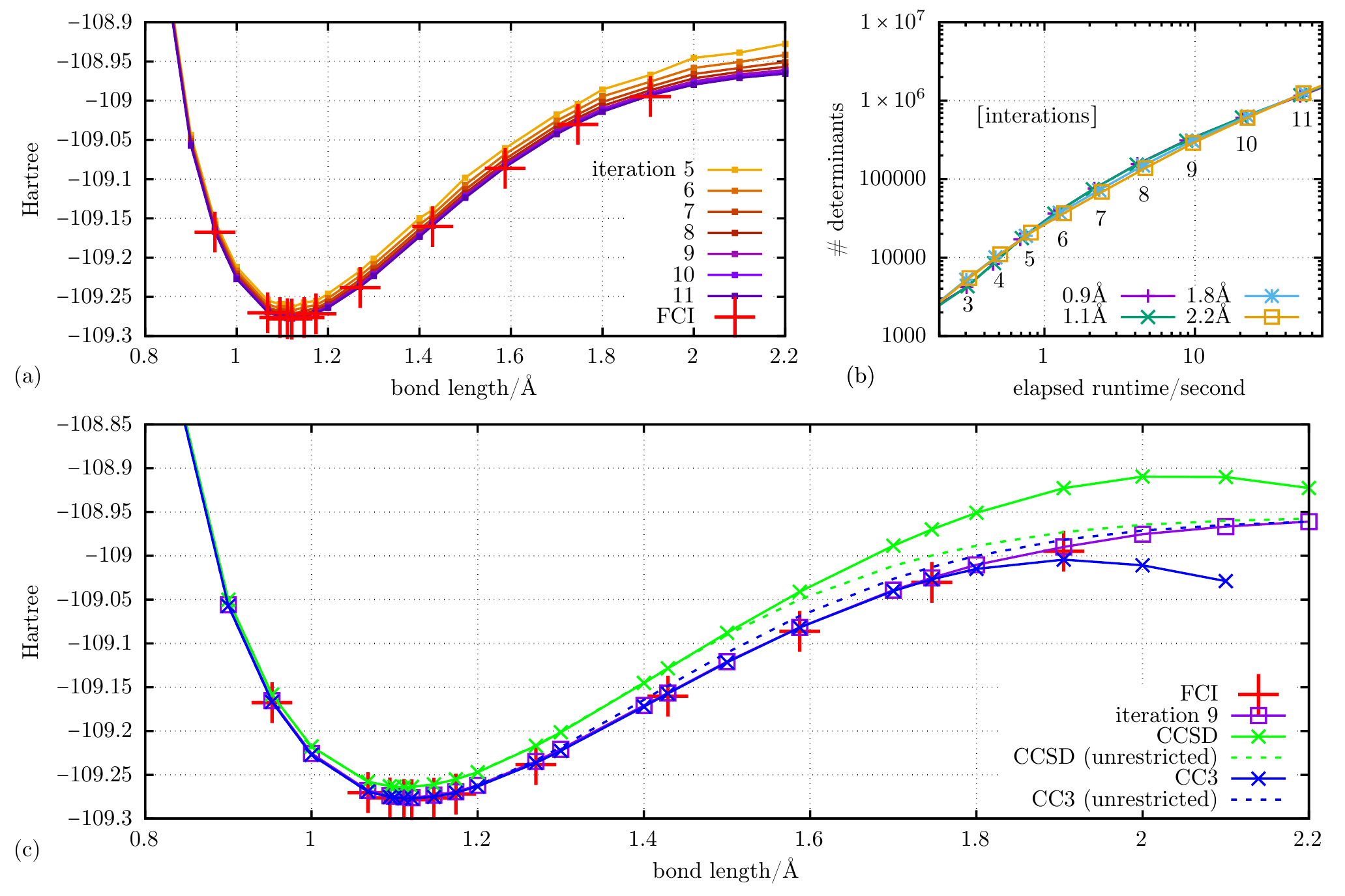}
    \caption{Treatment of N\sub2's full potential energy surface (PES) to near-dissociation in the cc-pVDZ basis (28 orbitals, 2 frozen): (a) treatment with 5--11 iterations of the present truncated Davidson approach plotted with benchmark FCI results taken from \Ref{LarsenEtAl2000}; (b) number of determinants (labeled with Davidson iteration number) versus elapsed runtime (\laptop) for 4 geometries (inner repulsive wall, near-equilibrium, bond-breaking, near-dissociation); (c) Davidson iteration~9 from panel~(a) along with FCI results compared with various non-perturbative coupled-cluster (CC) methods}
    \label{f:N2}
\end{figure*}

\subsection{N\sub2, full potential energy surface (PES)}\label{s:N2}

Presented next is a treatment of the nitrogen molecule in the double-$\zeta$ quality cc-pVDZ basis set with the 1s~cores of each N~atom frozen.
As in somewhat recent work developing the adaptive CI (ACI) method,\cite{SchriberEvangelista2016} here considered is the entire potential energy surface (PES) of N\sub 2, 
nearly 
to the dissociation limit. This involves the challenging task of modeling the breaking of a triple bond, which implicates both strong static and dynamic correlation, as noted in more recent work developing the new rank-reduced FCI (RR-FCI) approach,\cite{FalesEtAl2018} which also treated the N\sub 2 PES as a benchmarking example. 

\Figure{N2} plots results calculated with the present truncated Davidson approach over nearly the entire N\sub 2 PES, along with exact FCI results taken from \Ref{LarsenEtAl2000}, and coupled-cluster (CC) results, at the CCSD and CC3 levels of theory (without perturbation theory), calculated with the \verb!Psi4! quantum chemistry package.\cite{SmithEtAl2020} 
For the truncated Davidson calculations, the initial starting state, $\bpsiin$ (see \Fig{truncated Davidson}, line~2), was taken to be the solution from a small complete active space (CAS) calculation of 8 orbitals, again, excluding the 2 frozen core orbitals, and a \emph{variable} $\delta$-threshold was employed, $\delta \in [\delta_\text{max},\delta_\text{min}]  = [10^{-3},10^{-8}]$, as explained and justified in the sections that follow below. (This is the variable range that is ``hardcoded'' at line~14 of the algorithm shown in \Fig{truncated Davidson}.) 

To illustrate convergence of the present approach, energies obtained from Davidson iterations 5--11 are plotted in \Fig{N2}(a). Iteration~11 is in perfect visual agreement with the exact FCI results and it is numerically correct to chemical accuracy. However, very reasonable agreement is also obtained with somewhat fewer iterations, and panel~(a) illustrates the relative smoothness by which the iterations converge to the exact result over the full range of the PES, as well as the fact that rough qualitatively agreement with the converged result is obtained with even just 5--6 iterations.

Panel~(b) plots the scaling of the number of determinants at each Davidson iteration versus the elapsed runtime at four different representative geometries/bond-lengths.  One sees that for essentially all iterations, the data (timings and number of determinants) shows very little dependence on bond-length.
For a given geometry, the chemically accurate results at iteration~11 are obtained in about a minute (running on the same \processor\ laptop referenced in \Section{H2O}). On the other hand, a rough but qualitatively-reasonable  calculation---5 or 6 iterations in panel~(a)---can be computed in only about one second. 
The corresponding scaling of the number of determinants, going from rough to accurate, is from the tens of thousands to about a million.

To give a sense of the error incurred by a less than perfect calculation, panel~(c) compares the accuracy of the present approach over the full range of the PES with CC theory. The results of iteration~9 from panel~(a) are chosen as representative, which are seen in panel~(b) to take about 10 seconds to generate (per geometry).
Visually, the results from iteration~9 match the FCI results for all bond lengths, whereas, the results from CC~theory do not quite accomplish the same. 
The CCSD results are off by some margin at equilibrium, but the errors grow substantially as the bond is stretched, finally giving a spurious peak in the PES near the dissociation limit. The unphysical peak is a consequence of the reference Hartree-Fock (HF) determinant being generated by an restricted HF (RHF) calculation in a situation where the dissociation products are two radicals. Accordingly, panel~(c) also shows the CCSD result when the calculation is based upon an unrestricted (spin-polarized) HF (UHF) solution, in which case the correct dissociation limit is reached, but the errors in the bond-breaking region (and at equilibrium) remain. 
To investigate the remaining error, \verb!Psi4! was run at the higher CC3 level of theory. With an RHF reference, CC3 gave results in good visual agreement with FCI at equilibrium and through the bond-breaking region (and in agreement with 9~iterations of the present approach) but, as with CCSD, it gave spurious results near dissociation. Running CC3 based on a UHF reference again cured the problem at dissociation, but it caused a slight worsening of the result in the bond-breaking region, as can be seen from visual inspection panel~(c).

Thus, of these CC approaches, none gave uniformly correct results over the full range of the N\sub2 PES, though CC3 with an UHF reference does perform very well, only having significant error in the challenging bond-breaking region, there being off from FCI by about 17.5~mHartree (i.e., about 10 times chemical accuracy). 
Nevertheless, the present approach---even with RHF-optimized molecular orbitals, which are a poor choice in the dissociation limit---was seen to work robustly over the full PES, in true black-box fashion. The present method would presumably converge even faster and still more uniformly if it were run with UHF-optimized molecular orbitals, particularly in the near-dissociation limit, but that was not possible to do in this work. 
Of course, there are many other more sophisticated (and more expensive) flavors of CC theory such as multi-reference CC methods and methods including up to quadruple excitations which, for N\sub 2, do perform better\cite{LaidigEtAl1987,KroghOlsen2001,CooperKnowles2010} (although seemingly not to chemical accuracy over the full PES), but those plotted in \Figure{N2} are standard approaches that were readily available.
In contrast to the vast plethora of CC methods, which presumably may or may not work depending on geometry, bond order, RHF versus UHF reference, etc., what is shown here---treating the challenging (albeit simple) example of  N\sub 2---tends to indicate that the present method will predictably give at least qualitatively reasonable results, independent of geometry or reference state, even if absolute energies may be imprecise for low numbers of iterations.
This robustness is not surprising given that the approach is a straightforward approximation~to~FCI.

\subsection{C\sub 2, exploring the expansion vector growth factor and $\boldsymbol \delta$-thresholds}

The exotic bonding\cite{BoothEtAl2011,ShaikEtAl2012} of the carbon dimer has made it a popular benchmark\cite{BoothEtAl2011,Evangelista2014,Olivares-AmayaEtAl2015,TubmanLeeEtAl2016,HolmesTubmanEtAl2016a} for assessing the capabilities of quantum chemistry methods for treating strongly correlated systems.
Benchmark all-electron results for C\sub2 at its experimental equilibrium bond-length of 1.24253\AA\ are given in \Ref{Olivares-AmayaEtAl2015}, computed essentially exactly with density matrix re-normalization group (DMRG) methods for double-, triple-, and quadrouple-$\zeta$ basis sets (cc-pVDZ, cc-pVTZ, and cc-pVQZ), and each will be compared here.

\begin{figure}
    \centering
    \IG{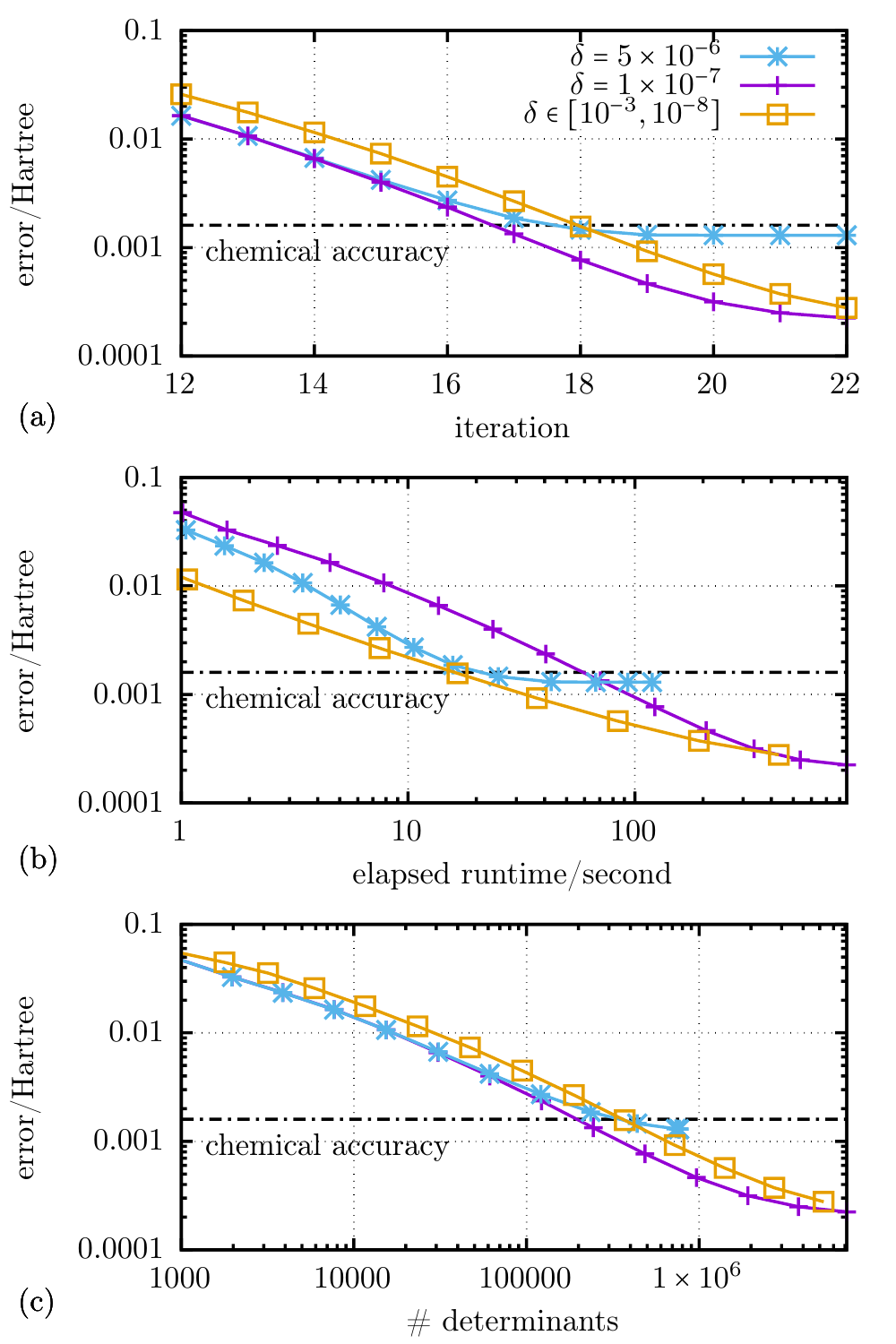}
    \caption{Convergence of the ground state energy of C\sub2 in the cc-pVDZ basis set (28~orbitals, none frozen) as computed with the present truncated Davidson approach at the experimental equilibrium bond length of 1.24253\AA\ using three different selections of the $\delta$-threshold:
    two calculations where $\delta$ is fixed and one where $\delta$ is allowed to vary at each iteration according to the prescription in \Fig{truncated Davidson} (line~14). The error, relative to the best DMRG result of $-75.731958$ Hartree from \Ref{Olivares-AmayaEtAl2015}, is plotted on a log-scale. (a) Error versus Davidson iteration count; (b) Error versus elapsed runtime (\laptop); (c) Error versus number of determinants}
    \label{f:C2,cc-pVDZ}
\end{figure}

For the presented truncated Davidson treatments of C\sub 2, the initial starting state, $\bpsiin$ (see \Fig{truncated Davidson}, line~2), was simply taken to be the Hartree-Fock solution. Depending on the experiment, the threshold parameter~$\delta$ was given either a fixed value (as done with the H\sub2O examples) or allowed to vary with iteration, as shown in \Fig{truncated Davidson} (line~14): i.e., computed as $\delta = 1/T$, where $T$ is the truncation size, but restricted to a range 
$[\delta_\text{max},\delta_\text{min}]$.  

\subsubsection{cc-pVDZ}

Calculations for C\sub 2 in the double-$\zeta$ cc-pVDZ basis set are now presented in order to illustrate (a)~the selection of the $\delta$-threshold and (b)~the benefit of growing the Davidson expansion vectors by a factor of 2 at each iteration. 

\paragraph{$\delta$-threshold selection.}

\Figure{C2,cc-pVDZ} illustrates the convergence behavior of the present approach for three different selections of the $\delta$-threshold: two calculations where $\delta$ is fixed and one where $\delta$ is allowed to vary at each iteration according to the prescription in \Fig{truncated Davidson} (line~14).
Panels (a)--(c) of \Fig{C2,cc-pVDZ} plot error (deviation from the best DMRG result from \Ref{Olivares-AmayaEtAl2015}) as a function of: (a)~Davidson iteration, (b)~elapsed runtime, and (c)~number of determinants.

Panel~(a) confirms that convergence to all accuracies requires the fewest iterations if one chooses the more-inclusive of the two fixed $\delta$-thresholds of $10^{-7}$, and that the calculation employing a variable $\delta$ is slowest to converge, though it about matches the less-inclusive $\delta$-threshold of $5 \times 10^{-6}$ in reaching chemical accuracy. 
However, if one poses the practical question of how much accuracy can be achieved for a given runtime, panel~(b) reveals that the variable $\delta$ calculation is actually superior for every runtime; and, noting that runtime is plotted on a log-scale, the difference is not insignificant, with the variable $\delta$ calculation reaching chemical accuracy in under 20~seconds (again, on the same \laptop).
Panel~(c) then confirms that the number of determinants is reasonably comparable between the three calculations, although the calculation with the variable $\delta$-threshold requires slightly more for the same accuracies, which is consistent with it being less accurate per iteration in panel~(a). 

\Figure{C2,cc-pVDZ} therefore illustrates, at least for this 28 orbital example, that varying $\delta$ is generally favorable from a runtime perspective (without any downsides). As an additional practical matter, though, the real advantage of the variable $\delta$ approach is that it mostly circumvents the question of how one chooses an appropriately-inclusive $\delta$-threshold for the problem at hand \textit{a priori}. The less-inclusive of the two $\delta$s in \Fig{C2,cc-pVDZ}, $\delta = 5\times 10^{-6}$, was specifically chosen, \textit{a posteriori}, because it was the least expensive value that was just barely sufficient to achieve chemical accuracy (see panel~(a)). On the other hand, the more-inclusive fixed $\delta = 1\times 10^{-7}$ provides a more reassuring level of convergence, but it is many times more expensive (in runtime) to reach chemical accuracy. Employing the variable $\delta$ selection procedure allows one to gain both benefits---speed and accuracy---without having to guess any parameters beforehand or, potentially worse, to have to run multiple calculations, start-to-finish, to verify convergence.  

\begin{figure}
    \centering
    \IG{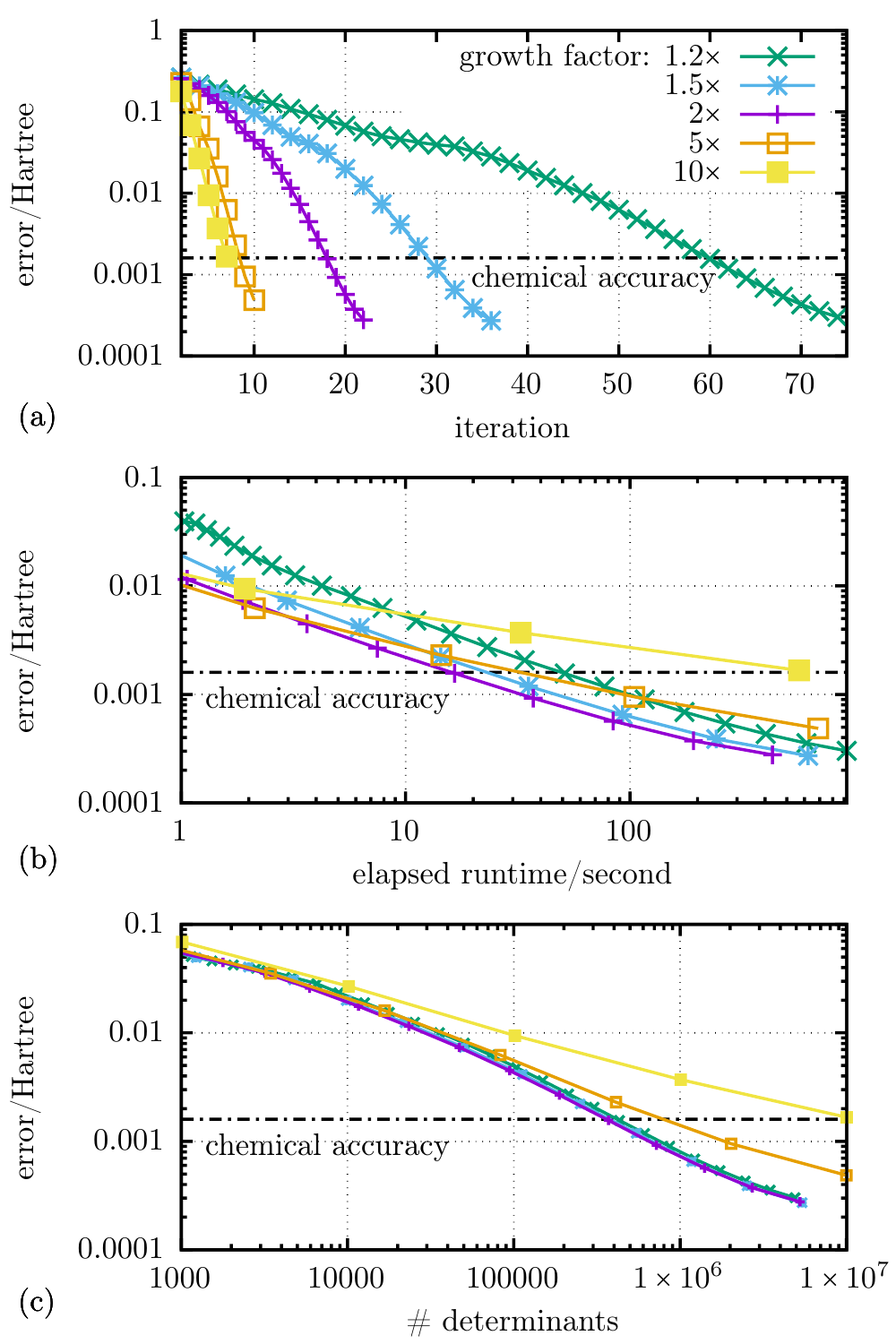}
    \caption{Convergence of the ground state energy of C\sub2 in the cc-pVDZ basis set (28 orbitals, none frozen) as computed with the present truncated Davidson approach at the experimental equilibrium bond length of 1.24253\AA\ using a sequence of different factors by which the Davidson expansion/basis vectors are expanded per iteration (e.g., $2 \times$ is hard-coded in \Fig{truncated Davidson} (line~13)). The error, relative to the best DMRG result of $-75.731958$ Hartree from \Ref{Olivares-AmayaEtAl2015}, is plotted on a log-scale. (a) Error versus Davidson iteration count; (b) Error versus elapsed runtime (\laptop); (c) Error versus number of determinants}
    \label{f:C2,cc-pVDZ,more}
\end{figure}

\paragraph{Size doubling at each iteration.}

\Figure{C2,cc-pVDZ,more} explores the advantages of size-doubling the Davidson expansion vectors at each iteration, i.e., using the growth factor of $2 \times$ which is ``hardcoded'' into the truncated Davidson method as presented in \Fig{truncated Davidson} (at line~13). 
Panels (a)--(c) of \Figure{C2,cc-pVDZ,more}---analogously to \Fig{C2,cc-pVDZ}---plot error (i.e., deviation from the DMRG result\cite{Olivares-AmayaEtAl2015}) versus (a)~Davidson iteration count, (b)~elapsed runtime, and (c)~number of determinants but, here, for a series of growth factors from a maximum of $10 \times$ down to only $1.2 \times$. 
Following the analysis of \Fig{C2,cc-pVDZ}, these calculations utilize the variable $\delta$-threshold, $[\delta_\text{max},\delta_\text{min}]  = [10^{-3},10^{-8}]$ but, again, it is the growth factor explored here (not $\delta$) which actually determines the size of expansion vectors.

Panel~(a) clearly illustrates the obvious fact that larger growth factors result in faster convergence per iteration, although the difference between factors of $5 \times$ and $10 \times$ (the largest shown) is not too significant. Panel~(b) reveals, however---analogously to \Fig{C2,cc-pVDZ}---that accuracy versus iteration count is not the same thing as accuracy versus runtime, with the calculation employing a factor of $10 \times$ being by far the slowest to converge (again, with runtime plotted on a log-scale). And, what panel~(b) further shows is that the most rapid convergence is indeed seen with the growth factor set to $2 \times$, with the factor of $1.5 \times$ being just slightly behind. Panel~(c) then confirms that there is not a substantial difference in the number of determinants required except for the largest $10 \times$ expansion, which is grossly inefficient, and the $5 \times$ expansion to a somewhat lesser extent, reaffirming the advantage of small growth factors. The results presented here for this single molecule and basis set are given in order to illustrate the foregoing principles, but it is to be understood that this author has found the advantages of size-doubling to be generally true across all simulations performed with the present approach and thus, as discussed, the factor of $2 \times$ is uniformly adopted for the remainder of this presentation.

\subsubsection{cc-pVTZ}

Now presented are somewhat more expensive calculations for C\sub 2 in a triple-$\zeta$ cc-pVTZ basis set of 60 orbitals (none frozen). These are done not only to illustrate the scaling of the method, but also to confirm the generality of the variable $\delta$-threshold selections made in the previous subsection.

\newcommand{\rome}{AMD EPYC 7742 Rome}

\begin{figure}
    \centering
    \IG{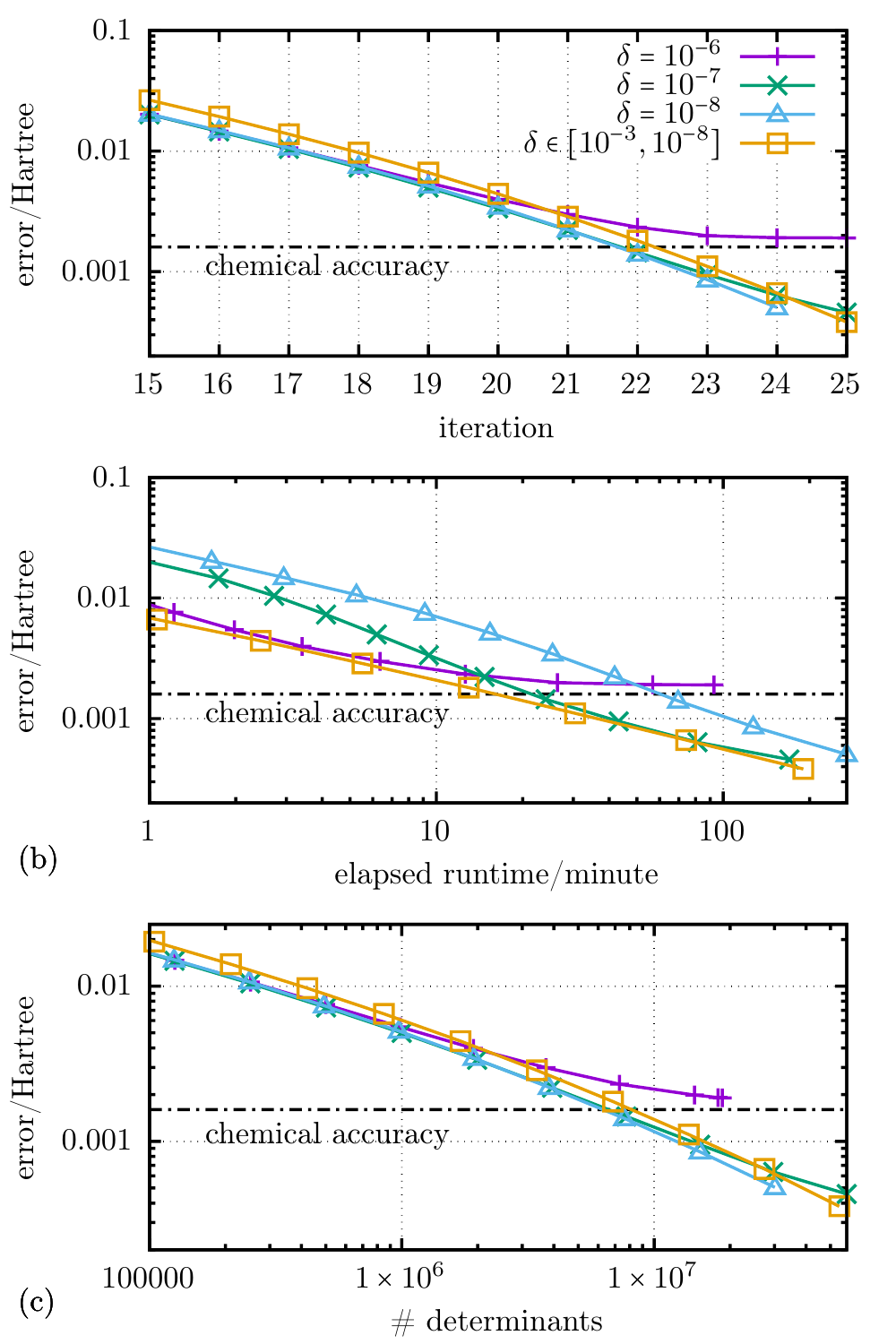}
    \caption{Convergence of the ground state energy of C\sub2 in the cc-pVTZ basis set (60 orbitals, none frozen) as computed with the present truncated Davidson approach at the experimental equilibrium bond length of 1.24253\AA\ using three different selections of the $\delta$-threshold:
    two calculations where $\delta$ is fixed and one where $\delta$ is allowed to vary at each iteration according to the prescription in \Fig{truncated Davidson} (line~14). The error, relative to the best DMRG result of $-75.809285$~Hartree from \Ref{Olivares-AmayaEtAl2015}, is plotted on a log-scale. (a) Error versus Davidson iteration count; (b) Error versus elapsed runtime (\laptop); (c) Error versus number of determinants}
    \label{f:C2,cc-pVTZ}
\end{figure}

\Figure{C2,cc-pVTZ} shows the results for \mbox{triple-$\zeta$} in three panels~(a)-(c) which are analogous to those shown in \Fig{C2,cc-pVDZ} for \mbox{double-$\zeta$}. Between the two figures, the results are generally consistent. 
One difference, however, shown in panel~(a) of \Fig{C2,cc-pVTZ}, is that, in this case, even a threshold of $\delta = 10^{-6}$ is not quite sufficient to achieve chemical accuracy, and that a more-inclusive $\delta$ of $10^{-7}$ or $10^{-8}$ is actually needed. But, panel~(a) does confirm the effectiveness of the variable $\delta$ approach from \Fig{C2,cc-pVDZ} and, as with the double-$\zeta$-sized calculation, panel~(b) shows the improved runtime efficiency one obtains, particularly at lower accuracies, as well as clear and consistent monotonic convergence, well past the target of chemical accuracy.
Likewise, panel~(c) shows that the absolute number of determinants does not vary greatly between using the fixed versus variable $\delta$s, except that if a fixed $\delta$-threshold is not inclusive enough to achieve the sought accuracy, at some point the number of determinants will increase significantly without benefit. 
Most importantly, panels~(a)--(c) confirm that the variable~$\delta$ approach circumvents the troubling question of how one chooses, \textit{a priori}, a $\delta$ small enough to achieve chemical accuracy (i.e., before one knows what the exact answer is) and large enough to not be unnecessarily burdensome, computationally. 

Finally, with regards to absolute runtimes, panel~(b) of \Fig{C2,cc-pVDZ} shows convergence to chemical accuracy in about 30~minutes on the same \laptop. 
This compares with convergence on the order of seconds for the double-$\zeta$ calculation of \Fig{C2,cc-pVDZ} but, of course, going from double- to triple-$\zeta$, the relative numbers of determinants in the full Hilbert space increases tremendously, here going from about 140 billion to about 2500 trillion, a difference of over 4 orders of magnitude.

\subsubsection{cc-pVQZ}

\begin{figure}
    \centering
    \IG{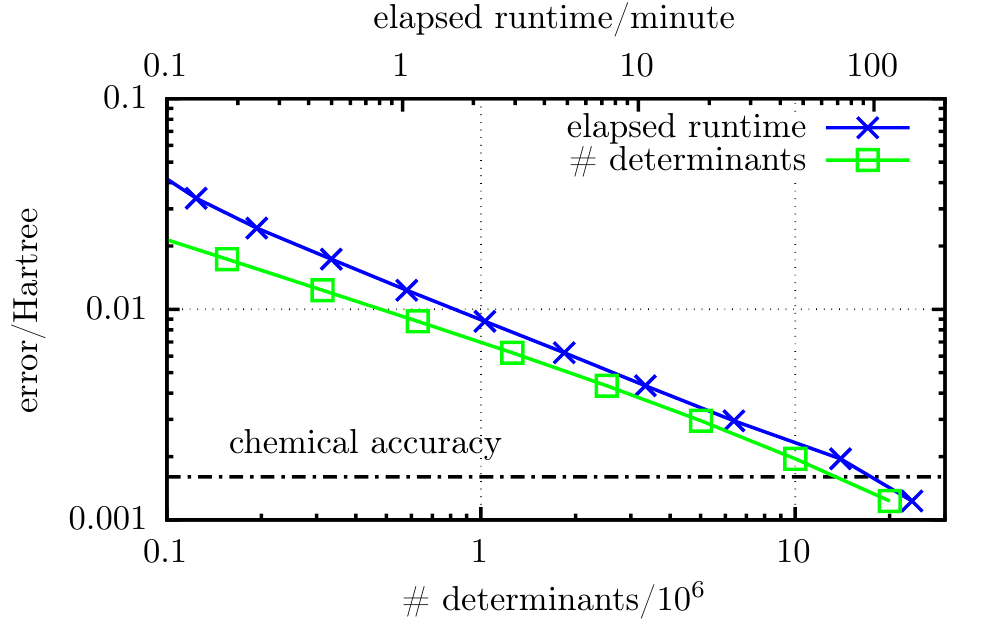}
    \caption{Convergence of the ground state energy of C\sub2 in the cc-pVQZ basis set (110 orbitals, none frozen) as computed with the present truncated Davidson approach at the experimental equilibrium bond length of 1.24253\AA. The error, relative to the best unextrapolated DMRG result of $-75.857231$~Hartree from \Ref{Olivares-AmayaEtAl2015}, is plotted on a log-scale versus number of determinants and also versus elapsed runtime (\laptop)}
    \label{f:C2,cc-pVQZ}
\end{figure}

As the final C\sub 2 example, calculations have been performed in the quadruple-$\zeta$ cc-pVQZ basis set of 110 orbitals (none frozen). Here, there are a whopping $4.6 \times 10^{18}$ determinants in the full Hilbert space. 
In this case, slightly improved efficiency was observed if the initial state was chosen to be the optimal wavefunction within the space of single- and double-excitations away from the Hartree-Fock state in a 8-electron, 8-orbital active space on top of frozen 1s~cores. In addition, due to the large number of orbitals, two iterations of natural orbital rotations were employed. With all other aspects of the algorithm remaining the same, this enabled a chemically-accurate calculation to be done in about 150 minutes on the same consumer-grade \laptop. 
\Fig{C2,cc-pVQZ} shows the results, again, illustrating smooth monotonic convergence, in this case, to a final wavefunction of about 20 million determinants.

\subsection{F\sub2: Benchmark comparison versus selective CI}

At this point in the discussion, it is interesting to further compare and contrast the present truncated full CI (FCI) approach with the general operational principles behind selective CI (SCI) approaches. 
The former clearly derives from subjecting the ubiquitous Davidson method to a progressive truncation scheme, whereas the latter, in its various forms,  are based on a variety of heuristic approaches for selecting a subset of determinants of the full Hilbert space, diagonalizing the Hamiltonian matrix in this space of determinants, typically with the Davidson (or a related) method, and then repeating the procedure by reselecting determinants and re-diagonalizing until convergence is achieved. 

On the surface, there seems to be little similarity between SCI and the standard Davidson approach of \Fig{Davidson}.
For one thing, the Davidson approach is essentially a way to diagonalize a large matrix, whereas large-matrix diagonalization is just one step, repeated at each iteration, of an SCI procedure. 
The relationship between the two becomes more apparent, however, when moving to the truncated approach of \Fig{truncated Davidson}. As described in \Section{theory}, avoiding full $\bsigma$-vector formation while maintaining the variational character of the Davidson approach implicates a distinct step of Hamiltonian matrix construction which involves finding connections between all the determinants arising in the expansion vectors. Furthermore, formation of the truncated expansion vectors themselves, in a sense, may be viewed as a determinant selection step because not all determinants are used, obviously, and, moreover, here, in generating the expansion vectors, the abbreviated Hamiltonian access described in \Section{theory} is the same strategy\cite{HolmesTubmanEtAl2016a} used in the variational part of the stochastic heat-bath CI (SHCI) approach.

Given the commonality with SHCI of abbreviated Hamiltonian operator access, and also the present use of techniques for Hamiltonian matrix construction which are modeled on, or at least inspired by, those used in the latest versions of SHCI,\cite{LiOttenEtAl2018} it is natural to briefly compare the performance of the present method to SHCI. This is perhaps most interesting because SHCI seems to be the well-developed and  best-performing of the many varieties of SCI,
and SCI, as a class of methods, seem to be amongst the most general, best performing techniques for approximating FCI results---see, e.g., the recent treatment of the benzene molecule (cc-pVDZ basis, frozen core) in~\Ref{EriksenEtAl2020}, with SHCI alongside many other methods. 

\begin{figure}
    \centering
    \IG{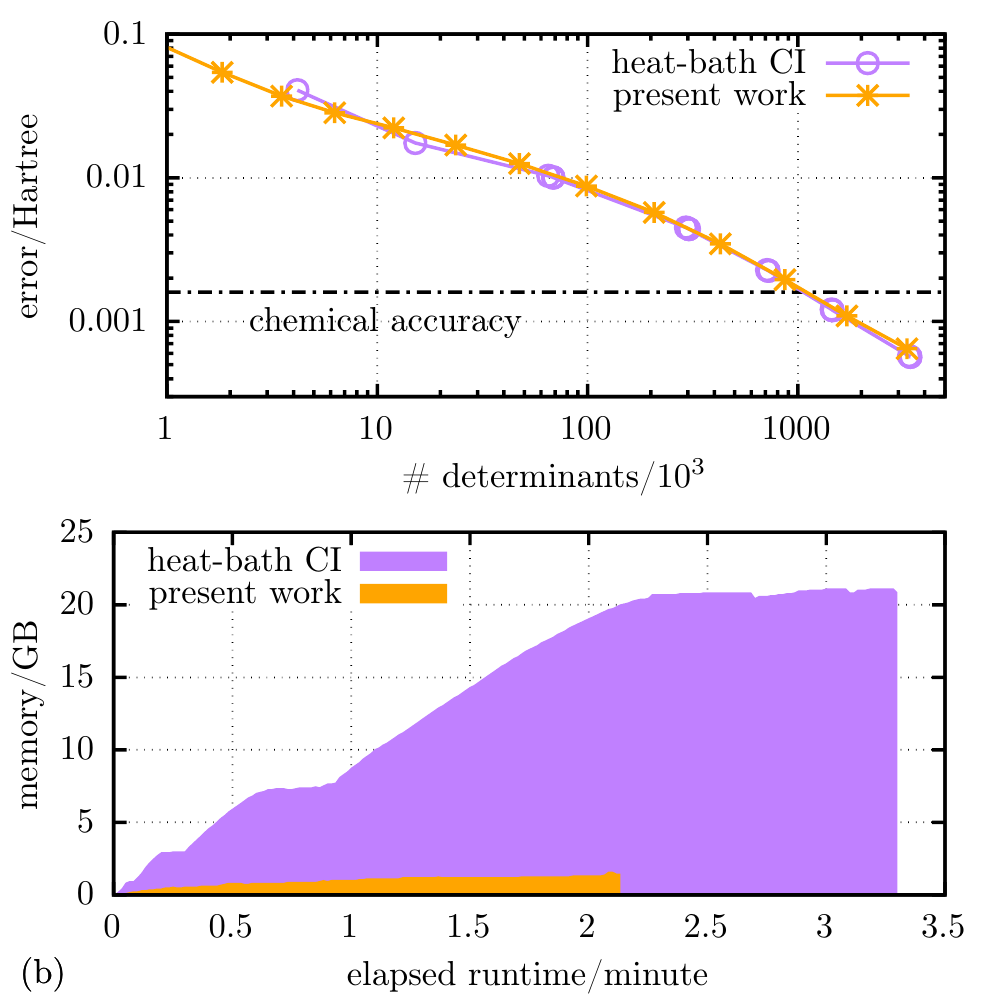}
    \caption{Performance comparison between the present truncated Davidson method and the variational component of stochastic heat-bath CI (SHCI) method for F\sub2 in the cc-pVDZ basis set (28 orbitals, 2 frozen) at the experimental equilibrium bond length of 1.4119\AA.
	    (a) Log-scale convergence of the ground state electronic energies given by both methods to within sub-mHartree accuracy of $-199.09941$~Hartree (the i-FCIQMC result from \Ref{ClelandEtAl2012}) versus number of determinants in the wavefunctions generated;
	    (b) Memory usage of both methods versus elapsed runtimes (\laptop) 
    }
    \label{f:F2}
\end{figure}

Accordingly, results are now presented for running the current approach alongside a recent version of the publicly-available \verb!Arrow! code,\footnote{%
\texttt{Arrow} is available at \texttt{https://github.com/QMC-Cornell/shci}. \texttt{Git} commit hash \texttt{054d5eaaa9e593c3e83ad27d854676492ea7a011}, the most recent of the \texttt{stable} branch, was compiled and used for the comparisons here} which implements the SHCI algorithms described in \Refs{HolmesTubmanEtAl2016a,SharmaHolmesEtAl2017a,LiOttenEtAl2018}.
The narrow objective is to assess whether the present truncated Davidson approach offers comparable performance with respect to the variational component of the SHCI algorithm (i.e., ignoring \verb!Arrow!'s impressive perturbative functionality), in the simple case of running on a single-processor/shared-memory architecture.
Thus, the system chosen for the comparison is F\sub 2 in a double-$\zeta$ cc-pVDZ basis of 28 orbitals with 2 frozen. 
F\sub 2 is the next homonuclear diatom beyond C\sub 2 and N\sub 2 with a singlet ground state, and the small basis allows easy treatment to chemical accuracy with both methodologies on the same Intel Core i7, 64 GB laptop used for the majority of the other examples treated in this work. 
The benchmark energy at the experimental bond-length of 1.4119\AA\ is taken from \Ref{ClelandEtAl2012} which employed the initiator full configuration interaction quantum Monte Carlo (i-FCIQMC) method.

\Figure{F2} shows the results. Panel~(a) demonstrates that both approaches converge to chemical accuracy and further to below 1~mHartree accuracy with about the same number of determinants, between about 1--4 million.
For this example, a standard input file from the \verb!Arrow! code was used that specified a schedule of progressively-inclusive determinant selection parameters,\footnote{The thresholds spec'd in the \texttt{Arrow} input file were: \\  \texttt{"eps\_vars": [ 5e-4,2e-4,1e-4,5e-5,2e-5 ], "eps\_vars\_schedule": [ 2e-3,1e-3,5e-4,2e-4,1e-4 ]}.} not unlike those chosen algorithmically in the present approach (here using the standard $[\delta_\text{max},\delta_\text{min}] = [10^{-3},10^{-8}]$, spec'd in \Fig{truncated Davidson}). The similarity is reflected in the similar convergence with the number of determinants shown in panel~(a).

\Figure{F2}(b) then compares computational performance, 
plotting memory usage in gigabytes (GB) versus elapsed runtimes.
The plot reveals that while both approaches have comparable runtimes,
memory usage of the code implementing the present truncated Davidson algorithm uses less than 8\% of the memory required by the \verb!Arrow! SHCI code, which is quite promising  considering that memory is generally the most critical resource when performing quantum mechanical calculations, and also given the relative maturity of the \verb!Arrow! code. One has to assume that this reduced memory footprint is reflective of what \emph{does} amount to a significant difference, in general, between the present truncated Davidson approach and SCI techniques: the fact that the present approach does not operate by repeatedly constructing, storing, and diagonalizing the large Hamiltonian matrix in the space of all selected determinants. And, one has to at least suspect that this cost savings will become much more significant as larger systems are treated via distributed-memory implementations where minimizing communication between nodes is critical to efficiency.

\subsection{Cr\sub 2: chemical accuracy from a fully variational treatment}

As a final example, now considered is the chromium dimer in the Alrichs~VDZ basis set\footnote{The Alrichs~VDZ basis set for the Cr-atom was downloaded from the Basis Set Exchange, \texttt{https://www.basissetexchange.org}} with the Mg core of each Cr frozen, leaving 24 electrons active in 30 orbitals, so at near half-filling. 
Although this basis set is apparently inadequate to describe Cr\sub 2's true electronic structure,\cite{LiYaoEtAl2020}
this example nevertheless appears frequently in the literature\cite{BoothEtAl2014,Olivares-AmayaEtAl2015,TubmanLeeEtAl2016,HolmesTubmanEtAl2016a,EriksenGauss2019,LiYaoEtAl2020,ZhangEtAl2020} as a standard benchmark
for testing methods of treating strong electron correlation.
The benchmark result for comparison here is again taken from \Ref{Olivares-AmayaEtAl2015} computed with DMRG for a bond-length of 1.5\AA, yielding a best unextrapolated energy of $-2086.420774$~Hartree.

\begin{figure}
    \centering
    \IG{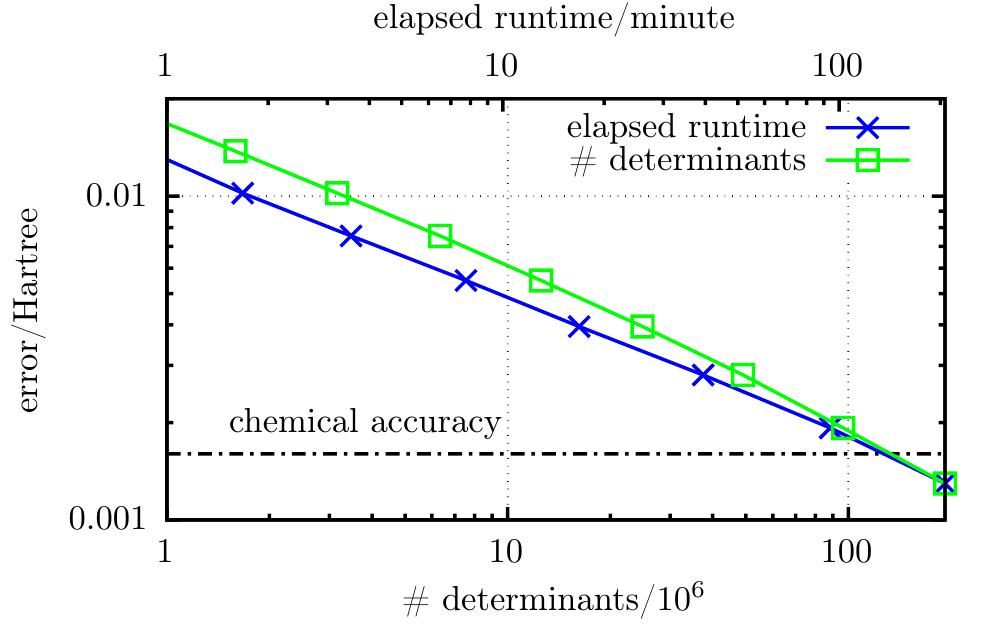}
    \caption{Convergence of the ground state energy of Cr\sub2 in the Alrichs~VDZ basis (42 orbitals, with 12 electrons of each Cr atom frozen, leaving an active space of 24 electrons in 30 orbitals) as computed with the present truncated Davidson approach, for the standard benchmark bond length of 1.5\AA.
    The error, relative to the best unextrapolated DMRG result from \Ref{Olivares-AmayaEtAl2015} of $-2086.420774$~Hartree, is plotted on a log-scale versus the number of determinants and also versus elapsed runtime (dual \rome\ processors)}
    \label{f:Cr2}
\end{figure}

\Figure{Cr2} shows the results for Cr\sub 2 
and what is again seen is smooth monotonic convergence to chemical accuracy with the present approach, as characteristic of every example treated in this work. 
For this particular problem, the \verb!PySCF! quantum chemistry package\cite{SunEtAl2018} was used to generate the molecular integrals because it gave a better RHF energy ($-2085.57297079$ Hartree) than \verb!Psi4! (which was used for all the other examples). 
The initial state was simply taken to be this RHF solution.
As with the quadruple-$\zeta$-treatment of C\sub2, a series of natural orbital rotations with progressively increasing numbers of determinants were performed to provide better single-particle basis for the final sequence of iterations.
The final chemically-accurate electronic wavefunction having about 200~million determinants(!)\footnote{Recall, no calculations presented herein impose any symmetry considerations on the molecular orbitals and resultant integrals and, as a result, the compactness of the electronic wavefunction has probably been sacrificed to some extent.} was computed in about 200~minutes on a pair of server-class \rome\ processors, each having 64 cores running at 2.25 GHz.
This illustrates the scale of problems that may, in principle, currently be tackled with the present approach, though future distributed-memory implementations of the algorithm are expected to have capabilities far superior to what has currently been developed.

\section{summary, conclusions, and future work}\label{s:conclusion}

This work develops a ``chemically accurate'' full configuration interaction (FCI) procedure which follows naturally from a sparse vector-based truncation of the standard Davidson method and the intuition that each Davidson expansion vector (diagonally-preconditioned residual) double in size at each iteration. 
Various aspects of the method are developed and demonstrated through the treatment of 5 simple systems, H\sub 2O, N\sub 2, C\sub 2, F\sub 2, and Cr\sub 2, at various levels of basis set complexity and cost, up through a quadruple-$\zeta$ treatment of C\sub 2.
Smooth monotonically variational convergence of ground state energies to within chemical accuracy is demonstrated for every treated example. 

Pedagogically, the present truncated Davidson approach is intellectually distinct from the popular paradigm of selective configuration interaction (SCI), instead, being more logically related to the work of Knowles, Handy, and Mitrushenkov, and also of Rolik, Szabados, and Surj\'an. In practice and procedurally, the present approach does turn out to be closely related to SCI, and the current implementation draws extensively from the variational component of the stochastic head-bath CI (SHCI) methodology, arguably, the most well-developed of the many current SCI varieties. Nevertheless, the practical difference between this work and SCI is significant: the present work totally avoids the repeated large-matrix diagonalizations which are the hallmark of SCI approaches. 

In sum, some advantages of this new approach are:~
(i)~straightforward theoretical basis with only the intuition of size-doubling at each iteration;
(ii)~smoothly monotonic convergence to chemical accuracy;
(iii)~lack of large-matrix diagonalization in contrast to SCI;
(iv)~lack of free/ambiguous parameters other than the threshold parameter $\delta$ which is, nevertheless, specified via an algorithmic prescription;
and
(v)~computational performance to chemical accuracy roughly on par with the best SCI approaches, runtime-wise, with the potential for much reduced memory requirements.
Anticipated future work will include a straightforward extension of the present method to the treatment of excited electronic states as well as the development of a large parallel distributed-memory implementation.  
In view of these points and future directions,
it seems plausible that the standard/ubiquitous Davidson method as used in the context of quantum chemical FCI calculations can be to a large extent augmented/supplanted by the present approach,
likely in circumstances where FCI is feasible though expensive, and more likely, in the many more circumstances where FCI is completely intractable due to memory limitations, particularly in basis sets where many digits of precision carry little physical meaning. 

In closing, it is noted that although FCI has realistically very few applications \textit{per~se} in quantum chemistry (beyond benchmarking), so-called complete active space (CAS) methods---which are just FCI in a reduced selection of orbitals---are essentially the only totally generic methods for simultaneously treating ground and excited electronic states of chemical systems at arbitrary geometries away from equilibrium. 
CAS-based approaches are therefore a widely used key ingredient in the dynamical simulation of electronically non-adiabatic chemical phenomena. 
It is in this areas where the present truncated Davidson method, as well as the many SCI approaches, could potentially have their greatest impact. 
 
Furthermore, from a quantum computing perspective, the present approach's explicit exponential growth of the wavefunction per iteration---through size-doubling of the sparsely-stored Davidson expansion vectors---provides a very clear incremental demonstration of when/where a classical hardware-based quantum computation becomes intractable (due to runtime and/or memory requirements), and also, therefore, a very strong clue as to situations where true quantum hardware-based advantage may be practical and significant.

\begin{acknowledgements}

I am grateful for support from the NASA Ames Research Center and from the DARPA Quantum Benchmarking program under interagency agreement IAA 8839, Annex 130.
I am also extremely grateful to Dr.~Eleanor Rieffel of the QuAIL Group at NASA Ames for her encouragement, advice, and support.
I also thank Prof.~Martin Head-Gordon in the Chemistry Department at UC~Berkeley for his reading of the manuscript and helpful comments. 

Resources supporting this work were provided by the NASA High-End Computing~(HEC) Program through the NASA Advanced Supercomputing~(NAS) Division at Ames Research Center.

\end{acknowledgements}

\vspace*{16pt}

\appendix

\bibliography{zotero}

\end{document}